\documentclass[12pt,usenames,dvipsnames]{article}

\usepackage{latexsym}
\usepackage{amssymb,amsfonts,amsmath}
\usepackage{graphicx} 
\usepackage{indentfirst}
\usepackage{bbm}
\usepackage{amssymb}
\usepackage{verbatim}
\usepackage{amsmath, amsthm,amssymb}
\usepackage{mathrsfs}
\usepackage[breaklinks]{hyperref}
\usepackage{amsfonts}
\usepackage{dsfont}
\usepackage{cite}
\usepackage{xcolor}
\usepackage{enumerate}
\usepackage{cleveref}
\parskip=\medskipamount

\newcommand {\cD}{{\cal D}}

\newcommand {\cF}{{\cal F}}

\newcommand {\cJ}{{\cal J}}

\newcommand {\cL}{{\cal L}}
\newcommand {\cM}{{\cal M}}
\newcommand {\cN}{{\cal N}}
\newcommand {\cO}{{\cal O}}

\newcommand {\cQ}{{\cal Q}}

\newcommand {\cS}{{\cal S}}

\newcommand {\cV}{{\cal V}}

\newcommand {\cX}{{\cal X}}

\newcommand {\cZ}{{\cal Z}}

\newcommand{\bL}{{\bf L}}

\newcommand{\bR}{{\bf R}}

\def\a{\alpha}

\def\b{\beta}
\def\c{\chi}
\def\d{\delta}

\def\f{\phi}
\def\g{\gamma}

\def\j{\psi}
\def\k{\kappa}
\def\l{\lambda}
\def\m{\mu}
\def\n{\nu}
\def\o{\omega}

\def\q{\theta}
\def\r{\rho}
\def\s{\sigma}
\def\t{\tau}

\def\x{\xi}
\def\z{\zeta}
\def\D{\Delta}
\def\F{\Phi}
\def\J{\Psi}
\def\L{\Lambda}
\def\O{\Omega}

\def\S{\Sigma}
\def\U{\Upsilon}

\def\tr{{\rm tr}}
\def\rd{{\rm d}}
\def\ri{{\rm i}}
\def\re{{\rm e}}

\def\rb{{\rm b}}
\def\ra{{\rm a}}
\def\rc{{\rm c}}

\newcommand{\ve}{\varepsilon}                            %new
\newcommand{\cDB}{{\bar\cD}}                            %new

\newcommand{\pa}{\partial}                           %new
\newcommand{\hf}{\frac12}
\newcommand{\vf}{\varphi}
\newcommand{\be}{\begin{equation}}
\newcommand{\ee}{\end{equation}}
\newcommand{\bea}{\begin{eqnarray}}
\newcommand{\eea}{\end{eqnarray}}
\newcommand{\non}{\nonumber}
\newcommand{\1}{{\underline{1}}}
\newcommand{\2}{{\underline{2}}}

\newcommand{\bm}[1]{\mbox{\boldmath$#1$}}

\def\double #1{#1{\hbox{\kern-2pt $#1$}}}

\newcommand{\qb}{{\bar{\theta}}}

\newif\ifdtup

\def\tr{{\rm tr}}
\newcommand{\Tr}{{\rm Tr}}
\def\hBox{\hat\square}

\newcommand{\bsubeq}{\begin{subequations}}
\newcommand{\esubeq}{\end{subequations}}
\newcommand{\bai}{{\bar i}}
\newcommand{\baj}{{\bar j}}
\newcommand{\bak}{{\bar k}}
\newcommand{\bal}{{\bar l}}

\newcommand{\bau}{{\bar 1}}
\newcommand{\bad}{{\bar 2}}

\newcommand{\rL}{{\rm L}}
\newcommand{\rR}{{\rm R}}

\newcommand{\eol}{\notag \\}
\newcommand{\veps}{\varepsilon}

\numberwithin{equation}{section}
\newcommand{\tsD}{{\nabla}}

\newcommand{\nrho}{{\veps}}

\newcommand{\sSU}{\mathsf{SU}}
\newcommand{\sSL}{\mathsf{SL}}

\newcommand{\sSO}{\mathsf{SO}}

\newcommand{\sU}{\mathsf{U}}

\begin{document}

\begin{titlepage}
\begin{flushright}
April, 2025 \\
Revised version: July, 2025
\end{flushright}
\vspace{5mm}

\begin{center}
{\Large \bf 
On three-dimensional $\cN=4$ supersymmetry: 
maximally \\[2mm] supersymmetric backgrounds and massive deformations}
\end{center}

\begin{center}

{\bf Sergei M. Kuzenko${}^a$, Emmanouil S. N. Raptakis${}^a$, Igor B. Samsonov${}^a$ \\ and Gabriele Tartaglino-Mazzucchelli${}^b$} \\
\vspace{5mm}

\footnotesize{
	${}^{a}${\it Department of Physics M013, The University of Western Australia,\\
		35 Stirling Highway, Perth W.A. 6009, Australia}}  
~\\
\vspace{2mm}
\footnotesize{
	${}^{b}${\it 
		School of Mathematics and Physics, University of Queensland,
		\\
		St Lucia, Brisbane, Queensland 4072, Australia}
}
\vspace{2mm}
~\\
Email: \texttt{sergei.kuzenko@uwa.edu.au, emmanouil.raptakis@uwa.edu.au, \\igor.samsonov@uwa.edu.au, g.tartaglino-mazzucchelli@uq.edu.au}\\
\vspace{2mm}

\end{center}

\begin{abstract}
\baselineskip=14pt
Using the $\mathsf{SO} ({\cal N})$ superspace formulation for $\cal N$-extended conformal supergravity in three dimensions, we derive all maximally supersymmetric backgrounds in the ${\cal N} =4$ case. The specific feature of this choice is that the so-called super Cotton tensor $X^{IJKL} = X^{[IJKL]}$, which exists for ${\cal N} \geq 4$, is equivalent to the scalar $X$ defined by $X^{IJKL} = \varepsilon^{IJKL} X$. This scalar may be used as a deformation parameter. In the family of $(p,q)$ anti-de Sitter (AdS) superspaces with $p+q=4$, $p\geq q$, it is known that $X\neq 0$ exists only if  $p=4$ and $q=0$. In general, the $(4,0)$ AdS superspaces are characterised by the structure group $\mathsf{SL}(2,{\mathbb R}) \times \mathsf{SO} (4)$ and their geometry is determined by two constant parameters, $S$ and $X$, of which the former determines the AdS curvature, while the $R$-symmetry curvature is determined by the parameters $(X+2S)$ and $(X-2S)$ in the left and right sectors of $\mathsf{SU}(2)_{\rm L} \times \mathsf{SU}(2)_{\rm R}$, respectively. Setting $S=0$ leads to  the so-called deformed ${\cal N}=4$ Minkowski superspace ${\mathbb M}^{3|8}_X$ introduced thirteen years ago. We use projective-superspace techniques to construct general 
interacting supersymmetric field theories in ${\mathbb M}^{3|8}_X$ and demonstrate that they originate as massive deformations of the following two families of ${\cal N} =4$  theories in standard Minkowski superspace  ${\mathbb M}^{3|8}$: (i)  $\cN=4$ superconformal field theories; and (ii) $\cN=4$ supersymmetric gauge theories in 
${\mathbb M}^{3|8}$ which are not superconformal but possess the $R$-symmetry group $\mathsf{SU}(2)_{\rm L} \times \mathsf{SU}(2)_{\rm R}$. Extensions of the theories in (ii) to ${\mathbb M}^{3|8}_X$ necessarily contain Chern-Simons terms at the component level. We also demonstrate the generation of topologically massive $\cN=4$ supersymmetric gauge theories from radiative corrections in the hypermultiplet sector.

\end{abstract}
\vspace{5mm}

\vfill

\vfill
\end{titlepage}

\newpage
\renewcommand{\thefootnote}{\arabic{footnote}}
\setcounter{footnote}{0}

\tableofcontents{}
\vspace{1cm}
\bigskip\hrule

\allowdisplaybreaks

\section{Introduction}

In the family of three-dimensional (3D) supersymmetric theories, those with $\mathcal{N}=4$ supersymmetry are particularly interesting. On the one hand, they enjoy non-perturbative constraints similar to the ones emerging in 4D, $\cN=2$ theories. On the other hand, they exhibit a potentially richer infrared structure characterized by a factorization into distinct Higgs and Coulomb branches, each with hyperk\"ahler geometry~\cite{Seiberg:1996nz, Kapustin:1999ha}. The presence of mirror symmetry~\cite{IS,Hanany:1996ie}, relating different ultraviolet gauge theory descriptions to the same infrared fixed point, reflects the duality structures inherent to these theories. Furthermore, 3D, $\mathcal{N}=4$ theories provide a fertile ground for studying protected sectors, exact operator algebras, and moduli space dynamics~\cite{Bullimore:2015lsa}.

It is well known that in 3D quantum field theories (QFTs), the quantum dynamics of matter coupled to gauge fields is closely linked to Chern-Simons (CS) terms.
 These terms are classically marginal and play a key role in the dynamics of gauge theories~\cite{Deser:1982vy}.
In general, quantum corrections can generate or shift CS terms, particularly via parity-violating fermion loops~\cite{Niemi:1983rq,Redlich1,Redlich2}. However, supersymmetry \cite{Nishino:1991sr, Ivanov:1991fn} imposes strong constraints on this mechanism. In theories with extended supersymmetry (e.g., $\mathcal{N} \geq 3$), the CS levels are typically protected from quantum corrections~\cite{Gaiotto:2007qi, Hosomichi:2008jd}. A notable example is the $\mathcal{N}=6$ ABJM theory~\cite{Aharony:2008ug}.

The story of 3D extended supersymmetry becomes even richer when the field theories are placed on curved backgrounds. 
The calculation of several observables by using localisation techniques naturally leads to the study of 3D SQFTs, for example, on the (squashed) three-sphere, on $S^2\times {S}^1$, and on other supersymmetric backgrounds. The formulation of supersymmetric field theory on curved backgrounds can be systematically achieved by using off-shell (conformal) supergravity and superspace techniques, and it is a rich avenue of research on its own on which we made some important contributions over a decade ago, see, e.g., \cite{KLT-M11,KT-M11,KLT-M12,BKTM,Butter:2013goa,Butter:2013rba,Kuzenko:2013vha,Kuzenko:2013uya,SamsonovSorokin,KTM14}.  

In the context of supersymmetric field theories on curved backgrounds, a special role was noticed to be played by supergravity theories with $\cN\geq4$ supersymmetry. The reason is technical, but the implications are quite unique and far reaching. For $\cN\geq 4$ conformal supergravity, the Weyl multiplet includes, as the covariant field with the lowest conformal weight, a Lorentz scalar $X^{IJKL}$ which is a completely antisymmetric tensor in its $\sSO(\cN)$ $R$-symmetry indices \cite{Butter:2013goa}. In superspace, a superfield with the same structure defines the super-Cotton tensor, which completely characterises all the superconformal multiplet of torsions and curvatures of a local superconformal geometry. In the case of Poincar\'e supergravity backgrounds, this tensor survives and is linked to peculiar features of $\cN\geq 4$ supergeometries, for instance, the failure of Dragon's theorem for superspace geometries \cite{KLT-M11,Cederwall:2011pu,Greitz:2011vh}. In the $\cN=4$ case it holds $X^{IJKL}=X\ve^{IJKL}$, with $\ve^{IJKL}$ 
the completely antisymmetric Levi-Civita tensor for $\sSO(4)$, and the scalar $X$ triggering interesting deformations that will be a main focus of our paper.

The simplest cases of  maximally supersymmetric curved backgrounds are the three-sphere ${S}^3$ (Euclidean supersymmetry) and anti-de Sitter (AdS) spacetime $\rm AdS_3$. Gauge theories possessing $\cN=4$ supersymmetry on these curved backgrounds have the interesting feature of requiring in general the presence of both Chern-Simons and Yang-Mills 
terms. This feature was first established in the ${S}^3$ case in \cite{HHL,SamsonovSorokin} and soon extended to $(4,0)$ AdS supersymmetry in \cite{KTM14}.\footnote{Similar aspects of superfield models on sphere $S^2$ were studied in Ref.~\cite{SamsonovSorokin-S2}.} It is known that 3D AdS supersymmetry is particularly rich.
The point is that $\cN$-extended AdS supersymmetry exists in several incarnations which are labelled by two non-negative integers $ p\geq q$ such that 
$p+q=\cN$. 
This is due to the fact that the 3D anti-de Sitter  group is reducible, 
$$\mathsf{SO}_0(2,2) \cong \Big( \mathsf{SL}(2, {\mathbb R}) \times \mathsf{SL}( 2, {\mathbb R}) \Big)/{\mathbb Z}_2~,$$ 
and so are its simplest supersymmetric extensions,  
${\mathsf{OSp}} (p|2; {\mathbb R} ) \times  {\mathsf{OSp}} (q|2; {\mathbb R} )$, 
which are known as $(p,q)$ AdS supergroups. This implies, in turn, that there are several versions of $\cN$-extended AdS supergravity \cite{AT},  known as the $(p,q)$ AdS supergravity theories. 

Within the superspace formulation for $\cN$-extended conformal supergravity \cite{HIPT,KLT-M11}, 
there exist maximally supersymmetric backgrounds,  
denoted ${\rm AdS}_{(3|p,q)}$, 
with the isometry supergroup ${\mathsf{OSp}} (p|2; {\mathbb R} ) \times  {\mathsf{OSp}} (q|2; {\mathbb R} )$, which 
were constructed in \cite{KLT-M12}. The specific feature of these $(p,q)$ AdS superspaces is that they are conformally flat and, for $\cN=p+q \geq 4$, $X^{IJKL}=0$.\footnote{A comprehensive analysis of conformally flat superspaces in diverse  dimensions is given in 
\cite{BILS}.} 
However, it turns out \cite{KLT-M12} that for $\cN\geq 4$ there also exist $(\cN,0)$ AdS superspaces with a non-vanishing covariantly constant super Cotton $X^{IJKL}$. These superspaces are not conformally flat, and their isometry supergroups differ from 
${\mathsf{OSp}} (\cN|2; {\mathbb R} ) \times  {\mathsf{SL}} (2, {\mathbb R} )$.
Restricting to the $\cN=4$ case, in general, the $(4,0)$ AdS superspace is characterised by the structure group $\mathsf{SL}(2,{\mathbb R}) \times \mathsf{SO} (4)$ and its geometry is determined by two constant parameters, $S$ and $X$, of which the former determines the AdS curvature, while the $R$-symmetry curvature is determined by the parameters $(X+2S)$ and $(X-2S)$ in the left and right sectors of $\mathsf{SU}(2)_{\rm L} \times \mathsf{SU}(2)_{\rm R}$, respectively. 
 If $X\neq 0$, the superspace is denoted ${\rm AdS}_{(3|4,0)}^X$.\footnote{Critical $(4,0)$ AdS superspaces with  $X = \pm 2S$ \cite{KLT-M12} are maxiamlly supersymmetric solutions in the minimal  topologically massive $\cN=4$ supergraviy constructed in \cite{Kuzenko:2016tfz}.}
General off-shell $(4,0)$ AdS supersymmetric nonlinear sigma models were constructed in 2012 \cite{BKTM}. Two years later, Ref. \cite{KTM14} described the off-shell $(4,0)$ supersymmetric Yang-Mills theories in $\rm AdS_3$.  

It was shown in \cite{KLT-M12} that starting from 
${\rm AdS}_{(3|4,0)}^X$ and setting $S= 0$ leads to a deformed $\cN=4$ Minkowski superspace denoted ${\mathbb M}^{3|8}_X$. Its geometry is determined by the following algebra of covariant derivatives  $\cD_A = (\cD_a, \cD^{i\bai}_\a)$: 
\bsubeq 
\label{N=4MSS}
\bea
\{\cD_\a^{i\bai},\cD_\b^{j\baj }\}&=&
2\ri\,\ve^{ij}\ve^{\bai \baj }\cD_{\a\b}
+\,{2\ri}\ve_{\a\b}X (\ve^{\bai \baj } \bL^{ij} - \ve^{ij} \bR^{\bai\baj})
~,
\\
{[}\cD_\a^{i\bai} , \cD_b{]}&=&
0~,\qquad \qquad
{[}\cD_a,\cD_b{]}=0~.
\eea
\esubeq
The isometry supergroup of ${\mathbb M}^{3|8}_X$ is a deformation of the $\cN=4$ super Poincar\'e group, and the corresponding superalgebra is a non-central extension of the three-dimensional $\cN=4$ Poincar\'e superalgebra. Such non-centrally extended superalgebras do not exist in four and higher  dimensions. Although their existence was pointed out by Nahm in 1978 \cite{Nahm}, they have appeared explicitly in various string- and field-theoretic applications only in the last 25 years 
\cite{Blau:2001ne,Itzhaki:2005tu,LM,Gomis:2008cv,Hosomichi:2008qk,Bergshoeff:2008ta}.
We emphasise once more that, from the point of view of extended conformal supergravity  in three dimensions \cite{KLT-M11}, the deformation parameter $X$ in \eqref{N=4MSS} is an expectation value of the super Cotton scalar. Unitary representations of the non-centrally extended $\cN=4$ Poincar\'e superalgebra have been studied, e.g., in \cite{Bergshoeff:2008ta}. In general, the presence of non-vanishing $X$ makes supermultiplets massive; there are no massless representations if $X\neq 0$ \cite{LM}.

General $\cN=4$ supersymmetric field theories in ${\mathbb M}^{3|8}_X$ can, in principle, be obtained from those in ${\rm AdS}_{(3|4,0)}^X$ by performing a careful $S\to 0$ limit. For the $\cN=4$ supersymmetric nonlinear sigma models this procedure has been carried out in \cite{BKTM}. In the present paper we extend the results of  \cite{BKTM} to the case of general models for self-interacting vector multiplets and supersymmetric Yang-Mills theories with non-centrally extended $\cN=4$ Poincar\'e supersymmetry. It should be pointed out that some examples of 3D theories with non-centrally extended $\cN=4$ Poincar\'e supersymmetry have been given
in \cite{LM,Bergshoeff:2008ta} and further discussions can be found in recent 
works \cite{KNBalasubramanian:2024bcr,Jiang:2024svn}.\footnote{In \cite{KNBalasubramanian:2024bcr,Jiang:2024svn}
the authors study an $X\ne0$ deformation in Minkowski spacetime, finding that 3D, $\mathcal{N}=4$ SCFTs lead to IR Topological QFTs with emergent Chern-Simons terms, constrained by mirror symmetry and anomaly matching. Interestingly, the authors also notice the equivalence between this deformation and the one referred as ``universal mass deformation'' in \cite{Cordova:2016xhm}.}
However, the story is clearly still incomplete. 

The present paper has three main objectives:
\begin{enumerate} 
\item 
Using the $\mathsf{SO} ({\cal N})$ superspace formulation for 3D, $\cal N$-extended conformal supergravity \cite{KLT-M11}, we will  derive all maximally supersymmetric backgrounds in the ${\cal N} =4$ case. We are specifically interested in those maximally supersymmetric backgrounds which allow for a non-vanishing value of the super Cotton scalar $X$. We will address this problem by using the same type of superspace analysis that has been used in other spacetime dimensions and different amounts of supersymmetry \cite{Ideas,Corfu,Kuzenko:2013uya,KNTM,Kuzenko:2020www}.\footnote{The study of off-shell supergravity used to formulate supersymmetric field theories in curved backgrounds is an old subject that goes back to the early days of supersymmetry. More recently, supersymmetry on curved spacetimes and supergravity became a subject of intensive study due to their application to localisation techniques. Besides our papers mentioned before, useful references for the reader are the 4D works \cite{Festuccia:2011ws, Butter:2015tra}, the 3D paper \cite{Closset:2012ru}, and the review articles \cite{Corfu,Dumitrescu:2016ltq}.} The idea of this approach, initiated in 
\cite{Ideas}, is to study restrictions on the curved superspace geometry which guarantee the existence of a maximal number of fermionic isometries (Killing spinors). In the case of 3D, $\cN=4$ supergravity, for instance, the maximal number of fermionic isometries is eight. Besides the ${\rm AdS}$ superspaces already described in \cite{KLT-M12}, we find geometries that are locally $\mathbb{R} \times S^2$, $\text{AdS}_2 \times \mathbb{R}$, and pp-wave spacetimes. All these cases possess non-trivial $ R$-symmetry curvatures and admit arbitrary deformations by $X$. Interestingly, we also identify one geometry which is completely characterised by $X$ only and possesses a positive constant Lorentz curvature --- a feature which we believe has not been observed before, as for instance in the analyses of \cite{Ideas,Corfu,Kuzenko:2013uya,KNTM,Kuzenko:2020www}.

\item Our next goal is to engineer interacting field theories with non-centrally extended $\cN=4$ Poincar\'e supersymmetry, that is supersymmetric models in  ${\mathbb M}^{3|8}_X$. For this we will apply the projective-superspace techniques developed earlier for three-dimensional  $\cN=4$ supergravity-matter systems  \cite{KLT-M11, KN14}, $\cN=4$ superconformal $\s$-models \cite{Kuzenko:2010rp}, and the field theories possessing $(p,q)$ AdS supersymmetry with $p+q=4$ 
\cite{BKTM,KTM14}. The deformation parameter $X$ leads to massive deformations of the following two families of ${\cal N} =4$  theories: (i)  $\cN=4$ superconformal field theories; and (ii) $\cN=4$ supersymmetric gauge theories which are not superconformal but possess the $R$-symmetry group $\mathsf{SU}(2)_{\rm L} \times \mathsf{SU}(2)_{\rm R}$. The latter family includes $\cN=4$ super Yang-Mills Chern-Simons theories. As part of our analysis, we show how to reduce manifest off-shell $\cN=4$ results to $\cN=2$ superfields and to component fields.

\item Finally, we will study the possibility to generate  $\cN=4$ super Yang-Mills Chern-Simons theories from radiative corrections in the hypermultiplet sector.

\end{enumerate}

This paper is organized as follows. In section 2, we review the superspace formulation of conformal supergravity developed in \cite{KLT-M11}, which is based on a structure group given ${\sSL}(2,\mathbb{R})\times {\sSO}(4)$.
Section 3 is devoted to the classification of maximally supersymmetric backgrounds of $\cN=4$ supergravity and the analysis of how the deformation driven by $X$ is implemented in cases beyond the ${\rm AdS}_3$ ones. From section 4 on, we will focus our attention on the deformed $\cN=4$ Minkowski superspace, which we denote by ${\mathbb M}^{3|8}_X$, and supersymmetric field theories defined on this background. In particular, section 4 analyses various geometrical aspects of ${\mathbb M}^{3|8}_X$, including the description of its isometries, and the reduction to centrally extended $\cN=2$ supersymmetry. Results in section 4 represent key ingredients to study manifestly off-shell $\cN=4$ supersymmetric multiplets and models constructed out of them. Section 5 is devoted to the definition of various covariant left and right projective multiplets on ${\mathbb M}^{3|8}_X$ as well as their reduction to $\cN=2$ superfields. In section 6, we show how to define various manifestly supersymmetric models with deformed $\cN=4$ supersymmetry. This includes left and right hypermultiplet supersymmetric nonlinear sigma models characterised by a hyperk\"ahler cone $\cM_{\rL} \times \cM_{\rR}$ target space geometry. A remarkable feature of deformed supersymmetry is that a non-trivial scalar potential, $V = \frac{1}{4} X^2 (K_\rL + K_\rR) $, is sourced by the deformation where $K_\rL$ and $K_\rR$ are the left and  right  K\"ahler potentials generated by the homothetic conformal Killing vectors of the associated product of hyperk\"ahler cones \cite{BKTM}.
In the simple case of a set of free hypermultiplets, the previous potential is simply associated with mass terms for the hypermultiplets. In section 6, we also construct various models for off-shell deformed $\cN=4$ Abelian vector multiplets. To conclude, in section 7 we describe super Yang-Mills couplings for the $\cN=4$ deformed multiplets formulated in terms of $\cN=2$ superfields and show how the topologically massive $\cN=4$ gauge theory emerges from radiative corrections in the deformed $\cN=4$ hypermultiplet models.
The main body of the paper is accompanied by three appendices. In Appendix \ref{3Dconventions}, we collect our notation and conventions, while in Appendix \ref{Cone} we review some properties of (hyper) K\"ahler cones. In Appendix \ref{ComponentSYM} we present the component field form of the $\cN=4$ super Yang-Mills Chern-Simons action.

%%%%%%%%%%%%%%%%%%%%%%%%%%%%%%%%%%

\section{$\cN=4$ supergravity in superspace}

Within  the geometric formulation for 3D, $\cN$-extended conformal supergravity sketched in \cite{HIPT} and fully developed in \cite{KLT-M11}, the $\cN=4$ structure group is ${\sSL}(2,\mathbb{R})\times {\sSO}(4)$. In order to define a large class of matter multiplets coupled to supergravity, it is advantageous to make use of the isomorphism 
${\sSO}(4) =  \big( {\sSU}(2)_{\rL}\times {\sSU}(2)_{\rR}\big)/{\mathbb Z}_2$.
We use the notation $\j_i$ and  $\chi_{\bar i}$ to denote the isospinors which transform under the defining representations of $\sSU(2)_{\rm L}$ and  $\sSU(2)_{\rm R}$, respectively.\footnote{Our conventions for spinor and isospinor indices are described in Appendix \ref{3Dconventions}.}

The $\cN=4$ covariant derivatives written in isospinor notation $\cD_A = (\cD_a, \cD^{i\bai}_\a)$ take the form
\bea
\cD_{A}&=&E_{A}{}^M \partial_M
+\hf \O_{A}{}^{\b \g} \cM_{\b \g}
+\F_A{}^{jk} \bL_{jk}
+\F_A{}^{\baj \bak} \bR_{\baj \bak}
~.
\eea
Here $E_A{}^M$ denotes the inverse supervielbein and is a non-degenerate supermatrix, $\O_A{}^{\b \g}$ is the Lorentz connection, and finally $\F_A{}^{j k}$ and $\F_{A}{}^{\baj \bak}$ denote the $\sSU(2)_\rL$ and $\sSU(2)_\rR$ connections, respectively. The two sets of $\sSU(2)$ generators act on the spinor covariant derivatives $\cD_\a^{i\bai}$ as follows:
\bea
&&
{\big [} {\bL}{}^{kl},\cD_{\a}^{i\bai}{\Big]} =\ve^{i(k} \cD_{\a}^{ l)\bai}
~,~~~
{\big [} {\bR}{}^{\bak\bal},\cD_{\a}^{i\bai}{\Big]} =\ve^{\bai(\bak} \cD_{\a}^{i \bal)}
~.
\label{acL-R}
\eea

Up to dimension-$1$, the algebra of covariant derivatives takes the form \cite{KLT-M11}
\bea
\{\cD_\a^{i\bai},\cD_\b^{j\baj }\}&=& 
2\ri\ve^{ij}\ve^{\bai \baj }(\g^c)_{\a\b}\cD_c
+{2\ri}\ve_{\a\b}\ve^{\bai \baj }(2\cS+X)\bL^{ij}
-2\ri\ve_{\a\b}\ve^{ij}\cS^{kl}{}^{\bai \baj }\bL_{kl}
+4\ri C_{\a\b}^{\bai \baj }\bL^{ij}
\non\\
&&
+2\ri\ve_{\a\b}\ve^{ij}(2\cS-X)\bR^{\bai \baj }
-2\ri\ve_{\a\b}\ve^{\bai \baj }\cS^{ij}{}^{\bak\bal}\bR_{\bak\bal}
+4\ri B_{\a\b}^{ij}\bR^{\bai \baj }
\non\\
&&
+2\ri\ve_{\a\b}(\ve^{\bai \baj }B^{\g\d}{}^{ij}+\ve^{ij}C^{\g\d}{}^{\bai \baj })\cM_{\g\d}
-4\ri(\cS^{ij}{}^{\bai \baj }+\ve^{ij}\ve^{\bai \baj }\cS)\cM_{\a\b}
~.
\label{N=4alg}
\eea
We see that the supergeometry is described in terms of two scalar fields $\cS$ and $X$, an iso-tensor $\cS^{ij \bai \baj}=\cS^{ji}{}^{\bai\baj}=\cS^{ij}{}^{\baj\bai}~.$ and the mixed tensors 
$B_{\a \b}^{i j}=B_{(\a \b)}^{(i j)}$ 
and $C_{\a \b}^{\bai \baj} =C_{(\a \b)}^{(\bai \baj )}$. It should be emphasised that $X$ is a supersymmetric extension of the Cotton tensor, hence a given background superspace is conformally flat if and only if $X = 0$, see \cite{Butter:2013goa} for the proof. 
Additionally, the torsion superfields in \eqref{N=4alg} are constrained to satisfy certain Bianchi identities, however the specific form of these constraints will not be required for the analysis undertaken in this work. Hence, we refer the reader to \cite{KLT-M11} for the technical details

An important property of the $\cN=4$ curved superspace geometry \eqref{N=4alg} is its invariance under 
the discrete transformation
\bea
\label{MirrorMap}
{\frak M} : ~ {\sSU}(2)_{\rL}~ \longleftrightarrow ~ {\sSU}(2)_{\rR}
\eea
which  changes the tensor types of superfields 
as 
${\rm D}^{(p/2)}_{\rL} \otimes {\rm D}^{(q/2)}_{\rR} 
\to {\rm D}^{(q/2)}_{\rL} \otimes {\rm D}^{(p/2)}_{\rR}$, 
where ${\rm D}^{(p/2)}$ denotes the spin-$p$ 
representation of $\sSU(2)$. In the rigid supersymmetric case, this transformation is 
an outer automorphism
of the $\cN = 4$ super-Poincar\'e algebra, which underlies mirror symmetry in  
3D $\cN=4$ Abelian gauge theories \cite{IS}.
It  has been  studied by Zupnik \cite{Zupnik3,Zupnik4} within the 3D $\cN=4$ rigid 
harmonic superspace 
\cite{Zupnik2}. Following \cite{Zupnik3,Zupnik4},
we  call $\frak M$ the {\it mirror map}. The torsion superfields in \eqref{N=4alg} transform as follows under \eqref{MirrorMap}:
\begin{subequations}
\label{mirror}
\begin{align}
	{\frak M}\cdot  \cS =&\cS ~,   &{\frak M}\cdot     \cS^{ij\bai\baj} = &\cS^{ij\bai\baj} ~,
	& {\frak M}\cdot X=&-X  ~, \\
    {\frak M}\cdot C_a^{\bai\baj}=&B_a^{ij} ~,  &{\frak M}\cdot B_a^{ij} =&C_a^{\bai\baj}~.
\end{align}
\end{subequations}

\section{Maximally supersymmetric backgrounds}

This section is devoted to the classification of maximally supersymmetric backgrounds of $\cN=4$ supergravity. We start by recalling the important result established in \cite{KNTM,Corfu}. Specifically, given a supergravity theory formulated in superspace, its maximally supersymmetric backgrounds are characterised by the following properties: (i) all Grassmann-odd components of the torsion vanish; and (ii) all Grassmann-even components of the torsion tensor are annihilated by the spinor covariant derivatives. In the present context this means that maximally supersymmetric backgrounds are characterised by the constraints:
\begin{subequations}
\label{MSBconstraints}
\begin{align}
		\cD_{\a}^{i\bai} \cS =& 0~, &\cD_{\a}^{i \bai} \cS^{jk\baj\bak} =& 0 ~, &\cD_{\a}^{i \bai} X =& 0~, \\
		\cD_{\a}^{i \bai} B_{\b \g}^{j k} = &0 ~, & \cD_{\a}^{i \bai} C_{\b \g}^{\baj \bak} =& 0~.
\end{align}
\end{subequations}

These constraints lead to significant simplifications in computing the algebra of covariant derivatives $\cD_A$. Specifically, we supplement Eq. \eqref{N=4alg} with the relations
\begin{subequations}
    \begin{align}
        [\cD_\a^{i \bai} , \cD_{\b \g} ] &= \Big \{  \ve_{\a (\b} \Big{[} 2 \d_{\g)}^\d \Big( \cS^{ij\bai\baj} + \ve^{i j} \ve^{\bai \baj} \cS \Big)  + \frac{4}{3} \Big( \ve^{\bai \baj} B_{\g)}{}^{ \d i j} + \ve^{i j} C_{\g)}{}^{ \d \bai \baj} \Big{)}  \Big{]}
        \non \\
        & \qquad
+        \d_{(\a}^{\d} \Big( \ve^{\bai \baj} B_{\b \g)}^{i j} + \ve^{i j} C_{\b \g)}^{\bai \baj} \Big)
        \Big \} \cD_{\d j \baj}~, \\
        [\cD_{\a \b} , \cD_{\g \d}] &= \ve_{\a (\g} B_{\d) \b}^{ij} (2\cS + X) \bL_{ij} + \ve_{\b(\g} B_{\d) \a}^{ij} (2\cS+X) \bL_{ij} \non \\
        &\phantom{=} + \ve_{\a (\g} C_{\d) \b}^{\bai \baj} (2\cS - X) \bR_{\bai \baj} + \ve_{\b(\g} C_{\d) \a}^{\bai \baj} (2\cS-X) \bR_{\bai \baj} \non \\
        &\phantom{=} + \ve_{\a(\g}\Big( B_{\d) \b}^{ij} B^{\l \r}_{ij} + C_{\d) \b}^{\bai \baj} C^{\l \r}_{\bai \baj} \Big) \cM_{\l \r} + \ve_{\b(\g}\Big( B_{\d) \a}^{ij} B^{\l \r}_{ij} + C_{\d) \a}^{\bai \baj} C^{\l \r}_{\bai \baj} \Big) \cM_{\l \r} \non \\
        &\phantom{=} - \Big( \cS^{ij\bai\baj} \cS_{ij\bai\baj} + 4 \cS^2 \Big) \Big(
        \ve_{\a (\g} \cM_{\d) \b} + \ve_{\b (\g} \cM_{\d) \a}
        \Big)~.
    \end{align}
\end{subequations}
Further, owing to the anti-commutation relations \eqref{N=4alg}, the conditions \eqref{MSBconstraints} have highly nontrivial implications. In particular, the integrability conditions arising from the fact that the anticommutator \eqref{N=4alg} acting on all the dimension-1 torsion superfields is zero, e.g. $\{\cD_\a^{i\underline{i}},\cD_\b^{j\underline{j}}\}\cS=0$, imply the following differential equations:
\begin{subequations}
	\label{DiffConstraints}
	\begin{align}
		&\cD_{\a \b} \cS = 0 ~, \qquad \cD_{\a \b} \cS^{i j \bai \baj} = 0~, \qquad \cD_{\a \b} X = 0~, \\
		&(\cD_{\a \b} - 2 \cS \cM_{\a \b}) B_{\g \d}^{i j} = 0 ~, \quad (\cD_{\a \b} - 2 \cS \cM_{\a \b}) C_{\g \d}^{\bai \baj} = 0~.
	\end{align}
\end{subequations}
These indicate that the torsion and curvature tensors are covariantly constant in the frame
\begin{align}
	\tilde{\cD}_A = (\tilde{\cD}_{\a}^{i \bai}, \tilde{\cD}_a) ~, \qquad \tilde{\cD}_{\a}^{i \bai} = \cD_{\a}^{i \bai} ~, \qquad \tilde{\cD}_{a} = \cD_a - 2 \cS \cM_{a}~.
\end{align}
Such a frame will be chosen below when considering supergeometries where either $B_{\a \b}^{i j}$ or $C_{\a \b}^{\bai \baj}$ are non-zero.
In addition to the differential equations \eqref{DiffConstraints}, there are several algebraic conditions following from Eq. \eqref{MSBconstraints} and the form of the algebra \eqref{N=4alg}. They are as follows:
\begin{subequations}
	\label{AlgConstraints}
	\begin{align}
 		B_{\a \b i j} \cS^{i j \bai \baj} = &0 ~, & C_{\a \b \bai \baj} \cS^{i j \bai \baj} = &0~,& B_{\a \b}{}^{(i}{}_k \cS^{j) k \bai \baj} + C_{\a \b}{}^{(\bai}{}_\bak \cS^{ij \baj) \bak} =& 0~, \\
 		B_{\a \b}{}^{(ij} \cS^{kl) \bai \baj} = &0~, & C_{\a \b}{}^{(\bai \baj} \cS^{i j \bak \bal)} = &0 ~, & B_{(\a}{}^{\g i j} C_{\b) \g}{}^{\bai \baj} + \cS^{ij\bak (\bai} C_{\a \b}{}^{\baj)}{}_{\bak} = &0~,\\
 		B^{\a \b i j} C_{\a \b}^{\bai \baj} = &0 ~, & B_{(\a \b}^{i j} C_{\g \d)}^{\bai \baj} = &0~, & B_{(\a}{}^{\g i j} C_{\b) \g}{}^{\bai \baj} - \cS^{k (i \bai \baj} B_{\a \b}{}^{j)}{}_k =& 0~,\\
        \span\span	B_{(\a}{}^{\g k(i} B_{\b) \g k}{}^{j)} + (2 \cS + X) B_{\a \b}^{ij} = &0~, & C_{(\a}{}^{\g \bak ( \bai} C_{\b) \g \bak}{}^{\baj)} + (2 \cS - X) C_{\a \b}^{\bai \baj} = &0~, \\
 	\span\span \cS^{k(i \bak (\bai} \cS^{j)}{}_k{}^{\baj)}{}_{\bak} + 2 \cS \cS^{i j \bai \baj}= &0~, & X \cS^{i j \bai \baj} = &0~. \label{XS}
	\end{align}
\end{subequations}

The above conditions describe highly constrained supergeometries. In particular, the latter condition in Eq. \eqref{XS} indicates that at least one of $X$ and $\cS^{ij \bai \baj}$ must vanish. Hence, the maximally supersymmetric backgrounds may be split into three families wherein: (i) $X=0$ and $\cS^{ij\bai\baj}=0$; (ii) $X\neq0$ and $\cS^{ij\bai\baj}=0$; and (iii) $X=0$ and $\cS^{ij\bai\baj}\neq0$. Below, we will examine each case in turn.

\subsection{Case (i): $X=0$ and $\cS^{ij\bai\baj} =0$}

To begin, we consider the simplest case in which $X=0$ and $\cS^{ij\bai\baj}=0$ hold. This choice reduces the integrability conditions \eqref{AlgConstraints} to:
\begin{align}
	\label{2.11}
	B_{\a \b}^{i j} C_{\g \d}^{\bai \baj} = 0~, \qquad
	B_{(\a}{}^{\g k(i} B_{\b) \g k}{}^{j)} + 2 \cS B_{\a \b}^{ij} = 0~, \qquad
	C_{(\a}{}^{\g \bak ( \bai} C_{\b) \g \bak}{}^{\baj)} + 2 \cS C_{\a \b}^{\bai \baj} = 0~.
\end{align}
The first equation indicates that at least one of $B_{\a \b}^{i j}$ or $C_{\a \b}^{\bai \baj}$ must vanish. We first consider the special case that $B_{\a \b}^{i j} = 0$ and $C_{\a \b}^{\bai \baj} = 0$. This implies that the geometry is the conformally flat $(4,0)$ AdS superspace \cite{KLT-M12}, which is controlled solely by $\cS$.

Next, we consider the branch $B_{\a \b}^{i j}\neq0$, noting that supergeometries with non-vanishing $C_{\a \b}^{\bai \baj}$ are related to the present ones by the mirror map \eqref{mirror}. Constraints \eqref{2.11} then have the solution:
\begin{subequations}
	\begin{align}
		(i):&~ \cS = 0 \quad \implies \quad B_{\a \b}^{i j} = B_{\a \b} B^{ij} ~, \quad B^{ij} B_{ij} = 2~, \\
		(ii):&~ \cS \neq 0 \quad \implies \quad 
                B_{\a \b}^{i j} = - 2 \cS \L_{\a \b,}{}^{\g \d} \d_{\g}^{k} \d_{\d}^l (R^{T})_{k l,}{}^{i j} ~, \label{3.7b}
	\end{align}
\end{subequations}
where $\L \in \sSO(2,1)$ and $R \in \sSO(3)$. Case $(i)$ describes three inequivalent supergeometries depending on the value of $B^2 = B^a B_a$. Specifically, for $B^2<0$, $B^2>0$ and $B^2=0$, locally the bosonic bodies of the superspaces are $\mathbb{R} \times S^2$, $\text{AdS}_2 \times \mathbb{R}$, and a pp-wave spacetime\cite{MFS}, respectively. 

The constraint \eqref{3.7b}, proves to have non-trivial consistency conditions. Specifically, by making use of the algebra of vector covariant derivatives $\tilde{\cD}_{\a \b}$
\begin{align}
\label{3.8}
	[\tilde{\cD}_{\a \b} , \tilde{\cD}_{\g \d} ] &= 4 \cS \ve_{\a (\g} \tilde{\cD}_{\d) \b} + 4 \cS \ve_{\b (\g} \tilde{\cD}_{\d) \a} 
         + 2 \cS \Big( \ve_{\a (\g} B_{\d) \b}^{ij} + \ve_{\b(\g} B_{\d) \a}^{ij} \Big) \bL_{ij} \non \\
	& \phantom{=} + \Big( \ve_{\a(\g} B_{\d) \b}^{ij} B^{\l \r}_{ij} + \ve_{\b(\g} B_{\d) \a}^{ij} B^{\l \r}_{ij} \Big) \cM_{\l \r} + 2 \cS^2 \Big(\ve_{\a (\g} \cM_{\d) \b} + \ve_{\b(\g} \cM_{\d) \b}\Big)~,
\end{align}
it may be shown that the constraint $\tilde{\cD}_{\a \b} B_{\g \d}^{i j} = 0$ has the non-trivial integrability condition
\begin{align}
    [\tilde{\cD}_{\a \b}, \tilde{\cD}_{\g \d}] B_{\l \r}^{i j} = 0 \quad \iff \quad \cS = 0~.
\end{align}
However, owing to Eq. \eqref{3.7b}, this implies $B_{\a \b}^{ij} = 0$. Hence, this supergeometry is inconsistent except for the trivial flat Minkowski superspace geometry.

\subsection{Case (ii): $X \neq 0$ and $\cS^{ij\bai\baj} = 0$}

In this subsection we will describe the maximally supersymmetric backgrounds characterised by non-zero super Cotton scalar $X$. This choice reduces the system of equations \eqref{AlgConstraints} to:
\begin{align}
	\label{2.5}
	B_{\a \b}^{i j} C_{\g \d}^{\bai \baj} = 0~, \quad  B_{(\a}{}^{\g k(i} B_{\b) \g k}{}^{j)} + (2 \cS + X) B_{\a \b}^{ij} = 0~, \quad
		C_{(\a}{}^{\g \bak ( \bai} C_{\b) \g \bak}{}^{\baj)} + (2 \cS - X) C_{\a \b}^{\bai \baj} = 0~,~
\end{align}
indicating that at least one of $B_{\a \b}^{ij}$ and $C_{\a \b}^{\bai \baj}$ vanish. Without loss of generality\footnote{We note that supergeometries characterised by $B_{\a \b}^{i j} = 0$ and $C_{\a \b}^{\bai \baj} \neq 0$ can be obtained from the present ones via the mirror map \eqref{mirror}.} we consider the branch $C_{\a \b}^{\bai \baj} = 0$. 

The solution to \eqref{2.5} is then:
\begin{subequations}
	\begin{align}
		(i):&~ 2 \cS + X = 0 \quad \implies \quad B_{\a \b}^{i j} = B_{\a \b} B^{ij} ~, \quad B^{ij} B_{ij} = 2~, \\
		(ii):&~ 2 \cS + X \neq 0 \quad \implies \quad B_{\a \b}^{i j} = - (2 \cS + X) \L_{\a \b,}{}^{\g \d} \d_{\g}^{k} \d_{\d}^l (R^{T})_{k l,}{}^{i j} ~, \label{3.11b}
	\end{align}
\end{subequations}
where $\L \in \sSO(2,1)$ and $R \in \sSO(3)$. In case $(i)$, the algebra of vector covariant derivatives $\tilde{\cD}_a$ reduces to
\begin{align}
	[\tilde{\cD}_{\a \b} , \tilde{\cD}_{\g \d} ] = 4 \cS \ve_{\a (\g} \tilde{\cD}_{\d) \b} + 4 \cS \ve_{\b (\g} \tilde{\cD}_{\d) \a}  - \Big( \ve_{\a(\g} B_{\d) \b} B^{\l \r} + \ve_{\b (\g} B_{\d) \a} B^{\l \r} \Big) \cM_{\l \r}~,
\end{align}
which yields three distinct backgrounds depending the choice of $B^2 = B^a B_a$. Specifically, for $B^2<0$, $B^2>0$ and $B^2=0$, the bosonic bodies of the superspaces are deformations of the $\mathbb{R} \times S^2$, $\text{AdS}_2 \times \mathbb{R}$, and pp-wave spacetimes \cite{MFS}, respectively. 

Next, in case $(ii)$, the algebra of covariant derivatives $\tilde{\cD}_a$ takes the form
\begin{align}
\label{3.12}
	[\tilde{\cD}_{\a \b} , \tilde{\cD}_{\g \d} ] &= 4 \cS \ve_{\a (\g} \tilde{\cD}_{\d) \b} + 4 \cS \ve_{\b (\g} \tilde{\cD}_{\d) \a} 
         + (2\cS+X) \Big( \ve_{\a (\g} B_{\d) \b}^{ij} + \ve_{\b(\g} B_{\d) \a}^{ij} \Big) \bL_{ij} \non \\
	& \phantom{=} + \Big( \ve_{\a(\g} B_{\d) \b}^{ij} B^{\l \r}_{ij} + \ve_{\b(\g} B_{\d) \a}^{ij} B^{\l \r}_{ij} \Big) \cM_{\l \r} + 2 \cS^2 \Big(\ve_{\a (\g} \cM_{\d) \b} + \ve_{\b(\g} \cM_{\d) \b}\Big)~.
\end{align}
This solution is characterised by the property that the $\sSU(2)_\rL$ curvature is aligned along $R^T$. Additionally, the covariant constancy condition $\tilde{\cD}_{\a \b} B^{ij}_{\g \d} = 0$ together with the structure of this algebra leads to the following integrability condition
\begin{align}
    [\tilde{\cD}_{\a \b}, \tilde{\cD}_{\g \d}] B_{\l \r}^{i j} = 0 \quad \iff \quad \cS = 0~,
\end{align}
indicating that the supergeometry is controlled by the super-Cotton scalar $X$. Thus, Eq. \eqref{3.12} reduces to
\begin{align}
    [{\cD}_{a} , {\cD}_{b} ] &= - X \ve_{a b c} B^{c i j} \bL_{ij} + X^2 \cM_{a b}~.
\end{align}
Remarkably, this result indicates that the spacetime manifold is non-conformally flat and characterised by a positive cosmological constant. As a consistency check, it may be shown that the covariant derivatives $\cD_A$ of this supergeometry satisfy the Jacobi identities
\be 
0 = (-1)^{\ve_A \ve_C } 
[ \cD_{A} , [ \cD_B , \cD_C \} \} ~+~\text{(two cycles)}~.
\ee
To the best of our knowledge, this exotic maximally supersymmetric $\cN=4$ geometry is novel compared to more standard solutions within the cases (i), (ii) and (iii), and will be studied in more detail elsewhere.

To complete our study of maximally supersymmetric backgrounds with non-zero $X$, we now fix both $B_{\a \b}^{ij} = 0$ and $C_{\a \b}^{\bai \baj} = 0$; the supergeometry is controlled by $\cS$ and the super-Cotton tensor $X$. When $\cS=0$, we obtain deformed Minkowski superspace ${\mathbb M}^{3|8}_X$ \cite{KLT-M12} which we will study in more detail in section \ref{deformed-Minkowski}, and for $\cS \neq 0$ the resulting supergeometry is $(4,0)$ AdS superspace with non-vanishing $X$ \cite{KLT-M12}.

\subsection{Case (iii): $X=0$ and $\cS^{ij\bai\baj} \neq 0$}

Having described all maximally supersymmetric backgrounds with non-vanishing super-Cotton tensor $X$ above, we now consider supergeometries characterised by $\cS^{ij\bai\baj} \neq 0$ and $X=0$. In this case, the constraints \eqref{AlgConstraints} reduce to:
\begin{align}
	\label{2.9}
	B_{\a \b}^{i j} = 0~, \qquad C_{\a \b}^{\bai \baj} = 0~, \qquad 
	\cS^{k(i \bak (\bai} \cS^{j)}{}_k{}^{\baj)}{}_{\bak} + 2 \cS \cS^{i j \bai \baj}= 0~,
\end{align}
hence the supergeometry is dictated by $\cS^{ij\bai\baj}$ and $\cS$. The latter constraint is solved by:
\begin{subequations}
	\begin{align}
		(i):&~ \cS = 0 \quad \implies \quad \cS^{ij\bai\baj} = \hat{\cS} l^{ij} r^{\bai \baj}~, \quad l^{ik} l_{kj} = \d^i_j~, \quad r^{\bai \bak} r_{\bak \baj} = \d^{\bai}_\baj~, \quad \hat{\cS} \in \mathbb{R}~, \\
		(ii):&~ \cS \neq 0 \quad \implies \quad \cS^{ij\bai\baj} = - \cS \Big( w^{i\bai} w^{j \baj} + w^{j\bai} w^{i \baj} \Big)~, \quad w^{i\bak} w_{j \bak} = \d^i_j~, \quad w^{k \bai} w_{\k \baj} = \d^\bai_\baj~.
	\end{align}
\end{subequations}
Cases (i) and (ii) correspond to the $(2,2)$ and $(3,1)$ AdS superspaces \cite{KLT-M12}, respectively.

%%%%%%%%%%%%%%%%%%%%%%%%%%%%%%%%%%%%%%%%%%

\section{Deformed $\cN=4$ Minkowski superspace}
\label{deformed-Minkowski}

This section is devoted to a detailed study of the isometries of the deformed $\cN=4$ Minkowski superspace ${\mathbb M}^{3|8}_X$ defined by the graded commutation relations \eqref{N=4MSS}.

Embedded in ${\mathbb M}^{3|8}_X$ is a centrally extended  $\cN=2$ Minkowski superspace ${\mathbb M}^{3|4}_X$ \cite{BKTM}. The latter is characterised by spinor covariant derivatives $ \cD_\a$ and  $\bar \cD_\a$ which are obtained by $\cN=2$ projection from the operators $\cD_\alpha := \cD_\alpha^{1 \bar 1}$ and $\bar\cD_\alpha := -\cD_\alpha^{2 \bar 2}$ obeying the algebra
\begin{subequations}
\label{N=2MSS}
\begin{align}
\{\cD_\alpha, \bar\cD_\beta\} &= -2\ri\, \cD_{\alpha \beta} - 2 \ri \, \veps_{\alpha \beta} X \hat \cZ~, \qquad
\{\cD_\alpha, \cD_\beta\} = 0~, \\
[\cD_\alpha, \cD_b] &= 0~, \qquad [\cD_a, \cD_b] = 0~.
\end{align}
\end{subequations}
Here we have introduced the central charge operator 
\bea
\hat \cZ := \bL^{1 2} - \bR^{\bar 1 \bar 2}~, \qquad 
{[}\hat \cZ,\cD_\a^{1\bau}{]}=
{[}\hat \cZ,(-\cD_\a^{2\bad}){]}=0 ~ \implies ~
{[}\hat \cZ,\cD_\a{]}=0 ~.
\label{central-N2}
\eea
Every supersymmetric field theory in ${\mathbb M}^{3|8}_X$ can be reformulated in 
${\mathbb M}^{3|4}_X$ with two supersymmetries being non-manifestly realised. Such a realisation has  been described in \cite{BKTM} for the most general $\cN=4$ supersymmetric nonlinear sigma model in ${\mathbb M}^{3|8}_X$. 

\subsection{Isometries of ${\mathbb M}^{3|8}_X$ }

The isometry supergroup  of ${\mathbb M}^{3|8}_X$ is generated by Killing supervector 
fields, $\x=\x^a\cD_a+\x^\a_{i\bai}\cD_\a^{i\bai}$, obeying the equation
\bea
0&=&
\Big{[}
\x
+\hf \L^{\g\d}\cM_{\g\d}
+\L^{kl}\bL_{kl}
+\L^{\bak\bal}\bR_{\bak\bal}
,
\cD_A
\Big{]}
~.
\label{sK-1}
\eea
This Killing equation is equivalent to
\bsubeq
\bea
\cD_\a^{ i\bai}\x_{\b\g}&=&
4\ri\ve_{\a(\b}\x_{\g)}^{i\bai}
~,
\label{sK-2-1}
\\
\cD_\a^{i\bai}\x_{\b}^{j\baj}&=&
\hf\L_{\a\b}\ve^{ij}\ve^{\bai\baj}
+\L^{ij}\ve^{\bai\baj}\ve_{\a\b}
+\L^{\bai\baj}\ve^{ij}\ve_{\a\b}
~,
\label{sK-2-2}
\\
\cD_\a^{i\bai} \L_{\b\g}&=& 0
~,
\label{sK-2-1b}
\\
\cD_\a^{i\bai}\L^{kl}&=&
-2\ri\ve^{i(k}\x_{\a}{}^{l)\bai}X
~,
\label{sK-2-3}
\\
\cD_\a^{i\bai}\L^{\bak\bal}&=&
2\ri\ve^{\bai(\bak}\x_{\a}{}^{i\bal)}X
~,
\label{sK-2-4}
\eea
\esubeq
and
\bsubeq \label{sK-2}
\bea
\cD_a\x_b&=&
\L_{ab}
~,
\label{sK-2-5}
\\
\cD_{a}\x^\b_{j\baj}
&=&0~, \quad
\cD_a\L^{\b\g}
= 0~, \quad
\cD_a\L^{kl}= 0 ~, \quad \cD_a\L^{\bak\bal}=0
~.
\label{sK-2-8}
\eea
\esubeq
An infinitesimal isometry transformation acts on a tensor superfield $\mathfrak T$ (with suppressed indices) by the rule 
\bea
\d_\x {\mathfrak T} = \Big( \x
+\hf \L^{\g\d}\cM_{\g\d}
+\L^{kl}\bL_{kl}
+\L^{\bak\bal}\bR_{\bak\bal}\Big) {\mathfrak T}~.
\label{IsometryGeneral}
\eea

%%%%%%%%%%%%%%%%%%%%%%%%%%%%%%%%%%%%%

\subsection{Reduction of the isometries of ${\mathbb M}^{3|8}_X$ to ${\mathbb M}^{3|4}_X$}
\label{SecReduction}

The existence of the embedding of ${\mathbb M}^{3|4}_X$ into ${\mathbb M}^{3|8}_X$ 
follows from  the fact that the covariant derivatives  $\cD_a$,
$\cD_\a^{1\bau}$ and $(-\cD_\a^{2\bad})$, together with the central charge $\hat{\cZ}$, form a closed subalgebra, see eqs. \eqref{N=2MSS} and \eqref{central-N2}.
Here we reduce the isometry transformations of ${\mathbb M}^{3|8}_X$ to 
${\mathbb M}^{3|4}_X$.

 For any $\cN=4$ tensor superfield  ${\mathfrak T}(x,\q_{\imath\bar{\jmath}})$,
we define its $\cN=2$ projection by
\bea
{\mathfrak T}|:={\mathfrak T}(x,\q_{\imath \bar{\jmath}})|_{\q_{1\bad}=\q_{2\bau}=0}~.
\label{N2red-1}
\eea
By definition,  ${\mathfrak T}|$  depends on the Grassmann coordinates
$\q^\mu:=\q^\mu_{1\bau}$ and their complex conjugates,  $\qb^\mu=\q^\mu_{2\bad}$.

Let us now consider a Killing vector field $\x$ of ${\mathbb M}^{3|8}_X$, as specified by eqs.~\eqref{sK-1}--\eqref{sK-2}. Following the notation of \cite{BKTM}, we introduce $\cN=2$ bar projections of the $\cN=4$ Killing parameters:
\bsubeq
\begin{gather}
\t^a:=\x^a|
~,~~~
\t^\a:=\x^\a_{\1\1}|~,~~~
\bar{\t}^\a=\x^\a_{\2\2}|~,~~~
t:=\ri(\L^{12}+\L^{\bau\bad})|=\overline{t}
~, ~~~t^{ab}:= \L^{ab}|~
;
\label{N2proj-20-1}
\\
\ve^\a:=-\x^\a_{1\bad}|~,~~~
\bar{\ve}^\a=\x^\a_{2\bau}|
~,~~~
\s:= \ri (\L^{12}-\L^{\bau\bad})|=\bar{\s}
~;
\label{N2proj-20-2-0}
\\
\bar{\L}_\rL:=\L^{11}|~,~~~
\L_\rL=\L^{22}|
~,~~~
\bar{\L}_\rR=\L^{\bau\bau}|~,~~~
\L_\rR=\L^{\bad\bad}|
~.~~~~~~
\label{N2proj-20-2}
\end{gather}
\esubeq
The scalar parameters $t$ and $\s$ correspond to the diagonal subgroup of
the $R$-symmetry group $\sSU(2)_{\rm L} \times \sSU(2)_{\rm R}$
\bea
2\L^{12}|\bL_{12}+2\L^{\bau\bad}|\bR_{\bau\bad} 
= \ri \, t \hat \cJ +\ri \, \s \hat \cZ~, \qquad 
\hat \cJ := \bL^{1 2} + \bR^{\bar 1 \bar 2}~,
\eea
with $[\hat \cJ, \hat \cZ]=0$. The $\sU(1)$ generator $\hat \cJ$ acts on the $\cN=2$ spinor covariant derivatives as follows:
\bea
{[} \hat \cJ,\cD_\a ]=\cD_\a
~,\qquad
{[} \hat \cJ, \bar \cD_\a {]}=- \bar \cD_\a~.
\eea

The parameters $(\t^a, \, \t^\a,\,\bar{\t}_\a,\, t^{\a\b} ,\, t )$
describe the infinitesimal isometries of ${\mathbb M}^{3|4}_X$.
Such an isometry is generated by a Killing supervector field 
$\t=\t^a\cD_a+\t^\a\cD_\a+\bar{\t}_\a\bar \cD^\a$ such that 
\bea
\Big{[}\t+\hf t^{bc}\cM_{bc} + \ri\, t \hat \cJ +\ri\, \s \hat \cZ,\cD_A\Big{]}=0~,
\label{2_0-Killing_iso-def}
\eea
for some parameters $t^{ab}$, $t$ and $\s$.
This can be easily proven by $\cN=2$ projection of the equations (\ref{sK-1})--(\ref{sK-2-4}). The parameter $\s$ generates central charge transformations. It may be shown, using  the equations \eqref{sK-1}--\eqref{sK-2}, that 
\begin{subequations}
\label{N2ts-const}
\bea
\cD_\a t &=&0~, \\
\cD_\a \s&=& 2X \t_\a~,\\
\cD_a \t_b&=&   t_{ab}  = - t_{ba}~,  \\
\cD_\a t_{bc} &=&0~,\\
\bar \cD_\a \t^\b &=&0~,\\
\cD_\a  \t^\b &=& \hf t_\a{}^\b  +\ri \d_\a{}^\b t ~,\\
\cD_\a \t^{\b \g} &=& -2\ri  (\d_\a{}^\b \bar \t^\g + \d_\a{}^\g \bar \t^\b) ~.
\eea
\end{subequations}

The remaining parameters
$(\veps^\alpha, \,\bar\veps_\alpha,  \,\L_L, \,\bar\L_L, \,\L_R, \,\bar\L_R)$ are 
associated with the remaining two supersymmetries
and the off-diagonal $R$-symmetry transformations. 
Their properties include the following: 
\begin{subequations}
\begin{align}
\bar \cD_\a \L_{\rm L} =&0~, 
& \cD_\a \L_{\rm L}= & -2\ri \,X \ve_\a~, \\
\bar \cD_\a \L_{\rm R} =&0~, 
& \cD_\a \L_{\rm R}=& -2\ri \,X \bar \ve_\a~,\\
\cD_\a \ve_\b =& \ve_{\a\b} \bar \L_{\rR} ~, & \cD_\a \bar \ve_{\b} =& - \ve_{\a\b} \bar \L_\rL~, \\
\bar \cD_\a \ve_\b =& \ve_{\a\b} \L_{\rL} ~,
& \bar \cD_\a \bar \ve_\b =& -\ve_{\a\b}  \L_{\rR} ~.
\end{align}
\end{subequations}
The $\hat \cZ$-charges of these parameters are
\begin{subequations}
\begin{align}
\hat \cZ \L_\rL &= - \L_\rL~,& \hat \cZ \ve^\a &= - \ve^\a~, \\
\hat \cZ \L_\rR &=  \L_\rR~,& \hat \cZ \bar \ve^\a &= \bar \ve^\a~.
\end{align}
\end{subequations}

In describing the transformation laws of $\cN=4$ left projective multiplets realised in terms of $\cN=2$ superfields, we follow the analysis of \cite{BKTM} and represent 
\bea
\label{id4.17}
\ve_\a = \bar \cD_\a \bar \r_\rL, \qquad \hat \cZ \bar \r_\rL = - \bar \r_\rL~, 
\qquad \cD_\a \bar \r_\rL=0~.
\eea
Choosing 
\bea
\bar \r_\rL = -\frac{\ri}{2X} \bar \L_\rR~,
 \qquad   \r_\rL = \frac{\ri}{2X}  \L_\rR~,
\eea
we obtain the following useful relations between these parameters
\be
\L_\rL = - \hf \bar \cD^2 \bar \r_\rL~, \qquad
\bar \L_\rL = - \hf  \cD^2 \r_\rL~,
\ee
where we have denoted $\cD^2 = \cD^\a \cD_\a$ and  
$\bar \cD^2 = \bar \cD_\a  \bar \cD^\a $.

In the right sector, we introduce  a chiral parameter $\r_\rR$ by the rule
\be
\ve_\a = - \cD_\a \r_\rR, \qquad \hat \cZ  \r_\rR = -  \r_\rR~, 
\qquad \bar  \cD_\a \r_\rR=0~.
\ee
Then choosing 
\be
\r_\rR = -\frac{\ri }{2X} \L_\rL~,\qquad
\bar\r_\rR = \frac{\ri }{2X} \bar\L_\rL~,
\ee
we derive the following relations:
\begin{align}
\label{id4.22}
\r_\rR &= \frac{\ri }{4X} \bar \cD^2 \bar \r_\rL~, & 
\r_\rL &= -\frac{\ri }{4X} \bar \cD^2 \bar \r_\rR~,\\
\L_\rR &= - \hf \bar \cD^2 \bar \r_\rR~, &
\bar \L_\rR &= - \hf  \cD^2 \r_\rR~.
\end{align}
The superfield parameters $\r_\rL$, $\r_\rR$ and their conjugates include, at the component level, the parameters corresponding to the two non-manifest supersymmetries and the off-diagonal $R$-symmetry transformations. 

%%%%%%%%%%%%%%%%%%%%%%%%%%%%%%%%

\section{Covariant projective multiplets on ${\mathbb M}^{3|8}_X$}

General field theories with non-centrally extended  $\cN=4$ Poincar\'e supersymmetry can be formulated in terms of covariant projective multiplets
 on ${\mathbb M}^{3|8}_X$. Such multiplets 
were proposed in \cite{KLT-M12} in the framework of $\cN=4$ supergravity.\footnote{
In the case of 4D $\cN=2$ rigid supersymmetry, the projective superspace approach was proposed in 
\cite{KLR, LR1, LR2} and further developed in \cite{G-RRWLvU}. This formalism is analogous to the harmonic superspace approach \cite{GIKOS,GIOS}, although the two setting have several important differences. The relationship between the {\it rigid} harmonic and projective superspace formulations is spelled out in \cite{K_double, Jain:2009aj, Kuzenko:2010bd, Butter:2012ta}.
} Since the structure group of ${\mathbb M}^{3|8}_X$ includes the $R$-symmetry group $\sSU(2)_\rL \times \sSU(2)_\rR$, it is natural to introduce left and right isotwistors, $v_\rL= (v^i)  \in {\mathbb C}^2 \setminus  \{0\}$ and  $v_\rR=(v^{\bai})  \in {\mathbb C}^2 \setminus  \{0\}$,
and use them to define two subsets of spinor covariant derivatives: 
\begin{subequations}
\bea
\cD_\a^{(1)\bai}&:=&v_i\cD_\a^{i\bai}~, \\ 
\cD_\a^{(\bau)i}&:=&v_\bai\cD_\a^{i\bai}~.
\eea
\end{subequations}
It follows from \eqref{N=4MSS} that 
\begin{subequations}
\begin{align}
\big\{ \cD_\a^{(1)\bai}\, , \, \cD_\b^{(1)\baj} \big\}=&~{2\,\ri\,}\ve_{\a\b}\ve^{\bai\baj}X\bL^{(2)}~,
& \bL^{(2)} :=& v_iv_j\bL^{ij} ~, \label{SpinorDer.a}\\
\big\{ \cD_\a^{(\bau)i} \, , \, \cD_\b^{(\bau) j} \big\} =&-  {2\,\ri\,}\ve_{\a\b}\ve^{i j}X\bR^{(\bar 2)}~,
& \bR^{(\bar 2)} :=& v_\bai v_\baj \bR^{\bai \baj} ~.
\end{align}
\end{subequations} 
These anti-commutation relations imply that one can introduce new off-shell multiplets that are defined in 
${\mathbb M}^{3|8}_X \times {\mathbb C}P^1_\rL$ and ${\mathbb M}^{3|8}_X \times {\mathbb C}P^1_\rR$, known as covariant left and right projective multiplets. 

%%%%%%%%%%%%%%%%%%%%%%%%%%%%%%%%%%%%%%%%%%

\subsection{Left projective multiplets on ${\mathbb M}^{3|8}_X$}
\label{Section5.1}

A left projective multiplet of weight $n$,
$Q_\rL^{(n)}(z^M,v^i)$, is defined to be a Lorentz and $\sSU(2)_\rR$ scalar
superfield\footnote{Covariant projective multiplets in $\cN=4$ supergravity and certain $\cN=4$ AdS superspaces are consistently defined only under the conditions they are described by Lorentz and $\sSU(2)_\rR$ scalar superfields. These conditions may be relaxed in the ${\mathbb M}^{3|8}_X$ case. Nevertheless we will only work with those projective multiplets which may be uplifted to an arbitrary supergravity background.}  
that lives on ${\mathbb M}^{3|8}_X$,
is holomorphic with respect to the isospinor variable $v^i$ on an open subset 
of ${\mathbb C}^2 \setminus  \{0\}$, 
and is characterised by the  conditions:
\begin{itemize}

\item[(i)] $Q_\rL^{(n)}$  is  a homogeneous function of $v$ 
of degree $n$,
\be
Q_\rL^{(n)}(z,c\,v_\rL)\,=\,c^n\,Q_\rL^{(n)}(z,v_\rL)~, 
\qquad c\in \mathbb{C}^* \equiv {\mathbb C} \setminus  \{0\}~.
\label{weight}
\ee

\item[(ii)]  
$Q_\rL^{(n)}$ transforms under the isometry supergroup 
of ${\mathbb M}^{3|8}_X$ as
\begin{subequations}\label{harmult1}
\bea
\d_\x Q_\rL^{(n)} 
&=& \Big( \x + \L^{ij} \bL_{ij} \Big) Q_\rL^{(n)} ~,   \\ 
\L^{ij} \bL_{ij}  Q_\rL^{(n)}&=& -\Big(\L^{(2)} {\bm \pa}_\rL^{(-2)} 
-n \, \L^{(0)}\Big) Q_\rL^{(n)} ~, \qquad 
{\bm \pa}_\rL^{(-2)} :=\frac{1}{(v_\rL,u_\rL)}u^{i}\frac{\pa}{\pa v^{i}}~,\\
\L^{(2)} &:=&\L^{ij}\, v_i v_j 
~,\qquad
\L^{(0)} :=\L^{ij}
\frac{v_i u_j }{(v_\rL,u_\rL)}~,
\qquad (v_\rL,u_\rL):=v^iu_i~.
\label{W2t3}
\eea 
\end{subequations}
Here $\x$ denotes an arbitrary Killing supervector field of ${\mathbb M}^{3|8}_X$, Eq.  (\ref{sK-1}),
  $\L^{ij}$ the associated $\sSU(2)_\rL$ 
 parameter, and $u_\rL = (u_{i})$ is an additional  isotwistor that is only constrained by $(v_\rL,u_\rL)\ne0$ and otherwise is completely arbitrary.

\item[(iii)] $Q_\rL^{(n)}$  obeys the analyticity constraint
\be
\cD^{(1)\bai}_{\a} Q_\rL^{(n)} = 0~.
\label{ana}
\ee  
\end{itemize}

Three comments are in order. 
Firstly, the homogeneity condition (\ref{weight}) and the analyticity constraint (\ref{ana}) 
are consistent with the interpretation that the isospinor
$ v^{i} \in {\mathbb C}^2 \setminus\{0\}$ is   defined modulo the equivalence relation
$ v^{i} \sim c\,v^{i}$,  with $c\in {\mathbb C}^*$.
Therefore, the projective multiplets live in ${\mathbb M}^{3|8}_X \times {\mathbb C}P^1_\rL$.
 
Secondly, by construction the superfield $Q_\rL^{(n)}$ is independent of the additional  isotwistor
$u_\rL = (u_{i})$.
On the other hand, the transformation law \eqref{harmult1} and the parameters \eqref{W2t3} involve $u_{i}$. Nevertheless, $\d_\x Q_\rL^{(n)}$ proves to be independent of $u_i$, 
\bea
{\bm \pa}_\rL^{(2)}  Q_\rL^{(n)}=0 ~\implies ~ 
 {\bm \pa}_\rL^{(2)} \delta_\x Q_\rL^{(n)} =0~,\qquad 
{\bm \pa}_\rL^{(2)} :=\frac{1}{(v_\rL,u_\rL)}v^{i}\frac{\pa}{\pa u^{i}}~.
\eea

Thirdly, the constraints \eqref{ana} are consistent with the algebra \eqref{SpinorDer.a} since 
\bea
\bL^{(2)} Q_\rL^{(n)}=0 ~\implies ~\big\{ \cD_\a^{(1)\bai}\, , \, \cD_\b^{(1)\baj} \big\}Q_\rL^{(n)}=0 ~.
\eea
 
The space of left projective multiplets is endowed with a real structure.
Given a   projective weight-$n$ multiplet $ Q^{(n)}_\rL (v^{i})$, 
its {\it smile conjugate},
$ \breve{Q}^{(n)}_\rL (v^{i})$, is defined by 
\bea
 Q^{(n)}_\rL (v^{i}) \longrightarrow  {\bar Q}^{(n)}_\rL ({\bar v}_i) 
  \longrightarrow  {\bar Q}^{(n)}_\rL \big({\bar v}_i \to -v_i  \big) =:\breve{Q}^{(n)}_\rL(v^{i})~,
\eea
with ${\bar Q}^{(n)}_\rL ({\bar v}_i)  :=\overline{ Q^{(n)}_\rL (v^{i} )}$
the complex conjugate of  $ Q^{(n)}_\rL (v^{i})$, and ${\bar v}_i$ the complex conjugate of 
$v^{i}$. It may be shown that $ \breve{Q}^{(n)}_\rL (v)$ is a left  projective weight-$n$ multiplet.
In particular, $ \breve{Q}^{(n)}_\rL (v)$
obeys the analyticity constraint $\cD^{(1)\bai}_{\a}  \breve{Q}^{(n)}_\rL =0$,
unlike the complex conjugate of $Q^{(n)}_\rL (v) $.
It may be also checked that 
\bea
\breve{ \breve{Q}}^{(n)}_\rL(v_\rL) =(-1)^n {Q}^{(n)}_\rL (v_\rL)~.
\eea
Therefore, if  $n$ is even, one can define real projective multiplets, 
 $\breve{Q}^{(2n)} = {Q}^{(2n)}$.
Note that geometrically, the smile-conjugation is a composition of the complex conjugation
with the antipodal map on the projective space ${\mathbb C}P^1_\rL$.

When dealing with projective multiplets, it suffices to work in the north chart of 
${\mathbb C}P^1_\rL$ where  $v^{1}\neq 0$. In this chart, 
it is natural to introduce an inhomogeneous complex variable $\z \in \mathbb C$ for ${\mathbb C}P^1_\rL$ by the rule $\z = v^2 / v^1$, 
\bea
 v^i = v^{1} \,(1, \z_\rL) ~, \qquad v_i = v^1 ( -\z_\rL, 1)~.
 \eea

Given a left projective weight-$n$ multiplet $ Q^{(n)}_\rL(z,v_\rL)$, we can associate with it 
a new superfield $ Q^{[n]}_\rL (z,\z_\rL )$ defined as 
\bea
Q^{(n)}_\rL(z,v_\rL)~\longrightarrow ~ Q^{[n]}_\rL(z,\z_\rL) \propto Q^{(n)}_\rL (z,v_\rL)~, \qquad 
\frac{\pa}{\pa \bar \z_\rL} Q^{[n]}_\rL =0 ~.
\eea
In terms of $ Q^{[n]}_\rL (z,\z_\rL )$, 
the analyticity condition \eqref{ana} takes the equivalent form
\bea
\cD_\a^{1\bad}Q_\rL^{[n]}=\frac{1}{\z_\rL} \cD_\a^{2\bad}Q_\rL^{[n]}~,\qquad
\cD_\a^{2\bau}Q_\rL^{[n]}=\z_\rL\cD_\a^{1\bau}Q_\rL^{[n]}
~.
\label{anaN2}
\eea
Thus, any spinor covariant derivative acting on a projective multiplet can always be expressed in terms of $\cD_\a^{1\overline{1}}$ and/or $\cD_\a^{2\overline{2}}$. This is particularly useful in reducing projective multiplets to $\cN=2$ superfields.

The explicit form of  $ Q^{[n]}_\rL(z,\z_\rL) $ for various projective multiplets will be given later on.
This superfield  can be represented by a Laurent series
\bea
Q^{[n]}_\rL (z,\z_\rL) =
\sum_{q}^{p} Q_k (z) \z_\rL^k~,\qquad -\infty \leq q <p \leq +\infty~,
\label{seriess}
\eea
with $Q_k(z)$ some {ordinary $\cN=4$ superfields}. Here $p$
and $q$ are {\it invariants} of the supersymmetry transformations.

We now consider several important types of projective multiplets.
Off-shell $\cN=4$ supersymmetric $\sigma$-models can be described by polar multiplets.
A {\it left arctic} weight-$n$ multiplet $\U_\rL^{(n)} (v_\rL)$ is defined to be 
holomorphic in the north chart
of $\mathbb CP^1_\rL$, and so it can be represented in the form 
\bea
\U_\rL^{(n)} ( v_\rL) &=&  (v^{1})^n\, \U_\rL^{[n]} ( \z_\rL) ~, \qquad 
\U_\rL^{ [n] } ( \z_\rL) = \sum_{k=0}^{\infty} \U_k  \z_\rL^k 
~.
\label{arctic1}
\eea
Its smile-conjugates is a {\it left antarctic} weight-$n$ multiplet $\breve{\U}_\rL^{(n)} (v_\rL) $,
 \bea
\breve{\U}_\rL^{(n)} (v_\rL) &=& 
(v^{2}  \big)^{n}\, \breve{\U}_\rL^{[n]}(\z_\rL) =
(v^{1})^n\z_\rL^{n}\, \breve{\U}_\rL^{[n]}(\z_\rL) ~, \qquad
\breve{\U}_\rL^{[n]}( \z_\rL) = \sum_{k=0}^{\infty}  {\bar \U}_k \,
\frac{(-1)^k}{\z_\rL^k}~.~~~
\label{antarctic1}
\eea
The superfields $\U_\rL^{[n]} ( \z_\rL)$ and $\breve{\U}_\rL^{[n]}(\z_\rL)$ constitute the so-called left polar weight-$n$ multiplet.

A real weight-$(2n)$ superfield $G_\rL^{(2n)} (v_\rL)$ is given by 
\begin{subequations} \label{RealWeight-2n}
\bea
G_\rL^{(2n)}(v_\rL)&=& (\ri \, v^1 v^2)^n G_\rL^{[2n]}(\z_\rL)=( v^1)^{2n} (\ri \,\z)^n G_\rL^{[2n]}(\z_\rL)~,\\
G_\rL^{[2n]}(\z_\rL) &=& 
\sum_{k=-p}^{p} G_k  \z_\rL^k~,
\qquad  {\bar G}_k = (-1)^k G_{-k} ~.
\eea
\end{subequations}
In the $p=\infty$ case, we are dealing with a tropical weight-$(2n)$ multiplet. 
Of special significance is the choice $n=0$, since the gauge prepotential of a left vector multiplet is a tropical weight-$0$ multiplet. In the $p<\infty$ case, we are dealing with a real left $\cO(2n)$ multiplet.
Equivalently it is described by a real  projective weight-$2n$ superfield $G^{(2n)}_\rL (v_\rL)$ 
of the form:
\bea
G^{(2n)}_\rL (v_\rL) &=& G^{i_1 \dots i_{2n}} v_{i_1} \dots v_{i_{2n}} 
=\breve{G}^{(2n)}_\rL (v_\rL ) ~.
\eea
The analyticity constraint (\ref{ana}) is equivalent to 
\bea
\cD_\a^{(j \bar j} G^{i_1 \dots i_{2n} )} =0~.
\eea
The reality condition $\breve{G}^{(2n)}  = {G}^{(2n)} $ is equivalent to 
\bea
\overline{ G^{i_1 \dots i_{2n}} } &=& G_{i_1 \dots i_{2n}}
=\ve_{i_1 j_1} \cdots \ve_{i_{2n} j_{2n} } G^{j_1 \dots j_{2n}} ~.
\eea

One may also consider complex $\cO(n)$ multiplets described by symmetric  rank-$n$ isospinors $\cQ^{i_1 \dots i_n} (z)$ constrained by 
\bea
\cD_\a^{(j \bar j} \cQ^{i_1 \dots i_{n} )} =0~.
\eea
Such multiplets are off-shell for $n>1$, while the $n=1$ case corresponds to the on-shell left Fayet-Sohnius hypermultiplet. 
Associated with  $\cQ^{i_1 \dots i_n} (z)$ is a left projective weight-$n$ multiplet defined by 
\bea
\cQ^{(n)}_\rL (v_\rL) &=& \cQ^{i_1 \dots i_{n}} v_{i_1} \dots v_{i_{n}} ~.
\eea

%%%%%%%%%%%%%%%%%%%%%%%%%%%%%%%%%%

\subsection{Right projective multiplets on ${\mathbb M}^{3|8}_X$}

The discussion in subsection \ref{Section5.1} naturally extends to the case of right projective multiplets. There are two  key differences. 
Firstly, while the conditions \eqref{weight} and \eqref{harmult1} are equivalent up to exchanging all $\sSU(2)_\rL$ objects with $\sSU(2)_\rR$ ones as well as labels $\rL$ with $\rR$, a right projective multiplet of weight $n$,
$Q_{\rR}^{(n)}(z^M,v^\bai)$, satisfies the 
analyticity conditions
\bea
\cD_\a^{(\bau)i}  Q_\rR^{(n)} = 0~.
\label{ana-right1}
\eea
Secondly, introducing an inhomogeneous complex variable $\z_\rR \in \mathbb C$ for ${\mathbb C}P^1_\rR$ by the rule 
\bea
 v^\bai = v^{\bar 1} \,(1, \z_\rR) ~, \qquad v_\bai = v^{\bar 1} ( -\z_\rR, 1)~
 \eea
and associating with   $ Q^{(n)}_\rR(z,v_\rR)$
a new superfield $ Q^{[n]}_\rR (z,\z_\rR )$ defined by 
\bea
Q^{(n)}_\rR(z,v_\rR)~\longrightarrow ~ Q^{[n]}_\rR(z,\z_\rR) \propto Q^{(n)}_\rR (z,v_\rR\textbf{})~, \qquad 
\frac{\pa}{\pa \bar \z_\rR} Q^{[n]}_\rR =0 ~,
\eea
the analyticity condition \eqref{ana-right1} is equivalent to
\bea
\cD_\a^{1\bad}Q_\rR^{[n]}=\z_\rR\cD_\a^{1\bau}Q_\rR^{[n]}~, \qquad
\cD_\a^{2\bau}Q_\rR^{[n]}=\frac{1}{\z_\rR} \cD_\a^{2\bad}Q_\rR^{[n]}
~.
\label{ana-right2}
\eea

%%%%%%%%%%%%%%%%%%%%%%%%%%%%%%%%%

\subsection{Reduction of left projective multiplets to ${\mathbb M}^{3|4}_X$}

The analyticity  conditions \eqref{anaN2}
tell us that the dependence of $Q^{[n]}_\rL (x,\q_{\imath {\bar{\jmath}}} , \z_\rL)$ on the Grassmann coordinates $\q^\m_{1\bar{2}}$ and $\q^\m_{2\bar{1}}$
is completely determined by its dependence on the other coordinates  $\q^\m_{1\bar{1}} $ and  $\q^\m_{2\bar{2}}$.
In other words, all the information about $Q_\rL^{[n]}(\z_\rL)$ 
is encoded in its $\cN=2$ projection $Q_\rL^{[n]}(\z_\rL)|$ defined by 
\bea
U|:=U(x,\q_{\imath \bar{\jmath}})|_{\q_{1\bad}=\q_{2\bau}=0}~,
\eea
for any $\cN=4$ tensor superfield  $U(x,\q_{\imath\bar{\jmath}})$.

Since the transformed superfield $\d_\x Q_\rL^{(n)} $, Eq. \eqref{harmult1}, is independent of $u_i$, 
we may choose
\bea
u_i =(1,0) ~\implies ~  (v_\rL,u_\rL) =v^1
\eea  
in the north chart. For brevity, in this section we will use the notation $\zeta := \zeta_\mathrm{L}$.

\subsubsection{Polar weight-$n$ multiplets} 

General off-shell sigma models are realised in terms of the polar multiplets
${\U}_\rL^{[n]}(z,\z)$ and ${\breve\U}_\rL^{[n]}(z,\z)$, 
so it is important to discuss these multiplets.

The isometry transformations for the polar weight-$n$ multiplet, reduced to ${\mathbb M}^{3|4}_X$, are
\bsubeq \label{eq_dUpL-aa}
\bea \label{eq_dUpL}
\d_\x \U_\rL^{[n]}|
&=& \Big{[} 
\t +\ri t\Big(\z\frac{\pa}{\pa \z}-\frac{n}{2}\Big)
+\z\bar{\ve}^\a \cD_\a
-\frac{1}{\z}\ve_\a \cDB^\a
+\ri\s\Big(\z\frac{\pa}{\pa \z}-\frac{n}{2}\Big)
\non\\
&&~
+\Big( \z \bar{\L}_\rL 
+\frac{1}{\z}{\L}_\rL\Big)\z\frac{\pa}{\pa \z}
-n\,\z\bar{\L}_\rL
 \Big{]} \U_\rL^{[n]}| 
 ~,~~~~~~
 \\  
\d_\x \breve{\U}_\rL^{[n]} |
&=& \Big{[} 
\t
+\ri t\Big(
\z\frac{\pa}{\pa \z}
+\frac{n}{2}\Big)
+\z\bar{\ve}^\a \cD_\a
-\frac{1}{\z}\ve_\a \cDB^\a
+\ri \s\Big(
\z\frac{\pa}{\pa \z}
+\frac{n}{2}\Big)
\non\\
&&~
+\Big(\z  \bar{\L}_\rL
+\frac{1}{\z}{\L}_\rL\Big)\z\frac{\pa}{\pa \z}
+n \frac{1}{\z}{\L}_\rL
 \Big{]}
 \breve{\U}_\rL^{[n]}  |
 ~.
\eea 
\esubeq

As an example, let us consider the polar weight-one hypermultiplet
\begin{subequations}
\begin{align}
\U^{[1]}_\rL(\z)| &= \sum_{n=0}^\infty \z^n \U_{n} = \F + \z \Sigma + \cdots~, 
\qquad \bar \cD_\a \F =0~, \quad \bar \cD^2 \S =0~, \label{Taylor}
\\
\breve{\U}^{[1]}_\rL(\z)| &= \sum_{n=0}^\infty (-\z)^{-n} \bar \U_{n} ~.
\end{align}
\end{subequations}
The component superfields $\F:=\U_{0}$ and $\S := \U_{1}$
are chiral and complex linear, respectively. They describe physical $\cN=2$ multiplets. 
The other Taylor coefficients in \eqref{Taylor},  $\U_2, \U_3, \dots$, are complex unconstrained $\cN=2$ superfields. They describe auxiliary $\cN=2$ multiplets. 

The physical superfields have the following $\hat \cZ $ charges: 
\bea
\hat \cZ \F = - \hf \F~, \qquad \hat \cZ  \S = \hf \S~.
\label{Zcharge}
\eea
By using the
transformation law (\ref{eq_dUpL})
one obtains the hidden off-shell supersymmetry transformations
\begin{subequations}
\begin{align}
\delta \F &= \ve^\alpha \bar \cD_\alpha \Sigma
+ \L_\rL \Sigma
	= - \frac{1}{2} \bar \cD^2 (\bar \rho_\rL \Sigma)~, \\
\delta \Sigma &= (\bar\nrho^\alpha \cD_\alpha 
-\bar \L_\rL) \F
	- \bar \cD_\alpha (\nrho^\alpha \U_{ 2}) 
	= \bar \cD_\a \Big\{ 
	\frac{\ri}{2X} \cD^\a (\bar \L_\rL \F) 
	-\nrho^\alpha \U_{ 2} \Big\} 
	~.
	\end{align}
\end{subequations}

\subsubsection{Real weight-$(2n)$ multiplets}
\label{SectO2}

We will also need the transformation for the real weight-$(2n)$ multiplet $G_\rL^{(2n)}(v_\rL)$. In the north chart, it is represented as in \eqref{RealWeight-2n}. From Eq.~(\ref{harmult1}) we find
\bea
\d_\x G_\rL^{[2n]} |
&=&\Big{[}
\t
+\ri t\z\frac{\pa}{\pa \z}
+\z\bar{\ve}^\a \cD_\a
-\frac{1}{\z}\ve_\a \cDB^\a
+\ri \s\z\frac{\pa}{\pa \z}
\non\\
&&~
+\Big(  \bar{\L}_\rL\z+ \L_\rL\frac{1}{\z} \Big)\z\frac{\pa}{\pa \z}
+ n\Big( \L_\rL\frac{1}{\z}- \bar{\L}_\rL\z \Big)
 \Big{]} G_\rL^{[2n]} |
 ~.
 \label{L2n}
\eea 

As an example, let us consider a left $\cO(2)$ multiplet. It is described by a superfield $W^{ij}=W^{ji} = \overline{W_{ij}} $
obeying the constraints\footnote{Such a superfield originates as a gauge-invariant field strength of a right vector multiplet.} 
\be
\cD_\a^{(i\bai}W^{kl)}=0~.
\label{WLanalyticity}
\ee
 Associated with $W^{ij}$ is the real  projective weight-2 multiplet $W_\rL^{(2)}(v) = W^{ij}v_iv_j$. 
\begin{subequations}
\bea
W^{(2)}_\rL(v)
&=&
\ri (v^{1})^2 \z W^{[2]}_\mathrm{L}(\z)~, \\
W^{[2]}_\mathrm{L}(\z)|
&=&
\frac{1}{\z}\varphi
+G
-\z \bar\varphi
~, \qquad \bar \cD_\a \varphi = 0 ~, \qquad \bar \cD^2 G=0~,
\eea
\end{subequations}
where we have introduced the $\cN=2$ superfields
\bea
\varphi:=-\ri \, W^{22}|~,\qquad
G:=2\ri \,W^{12}|~,\qquad
\bar \varphi = \ri \,W^{11}|
~.
\eea
The fact that  $\varphi$ is chiral and  $G=\bar G$ is a real linear superfield, follows from the analyticity constraint (\ref{WLanalyticity}).
 
Eq.~(\ref{L2n}) specifies the $\hat \cZ$-charges of $\varphi$ and $G$, 
\bea 
\hat \cZ \varphi = - \varphi ~, \qquad \hat \cZ G = 0~,
\eea
and allows us to find the non-manifest supersymmetry transformation which relates these superfields:
\bsubeq 
\bea
\d \varphi
&=&
\big(\ve^\a\cDB_\a +\L_\rL\big) G =  - \frac{1}{2} \bar \cD^2 (\bar \rho_\rL G)
~, 
\\
\d G &=& \big(\bar{\ve}^\a\cD_\a - 2\bar{\L}_\rL\big)\varphi
- \big(\ve^\a\cDB_\a +2\L_\rL\big)\bar \varphi \non \\
&=&  \cD^\a \big (\bar{\ve}_\a\varphi)
+ \cDB_\a(\ve^\a\bar \varphi \big)
= \cD^\a \big(\varphi \cD_\a \r_\rL \big)+ \cDB_\a \big(\bar \varphi \bar \cD^\a \bar \r_\rL \big) \\
&=&  -  \cD^\a \bar \cD_\a  \big(\bar \r_\rR \varphi - \r_\rR \bar \varphi \big) ~.  \non 
\eea
\esubeq

%%%%%%%%%%%%%%%%%%%%%%%%%%%%%%%%%%%

\subsubsection{On-shell Fayet-Sohnius hypermultiplet}
\label{FShyper}

The on-shell left Fayet-Sohnius hypermultiplet is described by an isospinor superfield $q^i (z) $ and its conjugate  $\bar q_i (z)$. The former is constrained by
\bea
\cD_\a^{(i \bar i} q^{j )} =0~.
\label{FSconstraints}
\eea
Associated with $q^i$ are $\cN=2$ chiral $Q_+$ and antichiral $\bar Q_-$ superfields 
\bea
Q_+ := q^2|~, \qquad \bar \cD_\a Q_+ = 0~; \qquad 
\bar Q_- := q^1|~, \qquad \cD_\a \bar Q_- =0
~,
\eea
with the following  $\hat \cZ$ charges 
\bea
\hat \cZ Q_+ = - \hf Q_+~, \qquad \hat \cZ \bar Q_- = \hf \bar Q_-~.
\eea
The hidden supersymmetry transformation acts on them by the rule
\bea
\d Q_+ = \frac{1}{2} \bar \cD^2 (\bar \rho_\rL \bar Q_- )~, \qquad 
\d \bar Q_- = -\frac{1}{2} \cD^2 ( \rho_\rL Q_+)~,
\label{FS-SUSY}
\eea
where we have used the equations of motion $\bar \cD^2 \bar Q_- =0$ and $\cD^2 Q_+ =0$, which follow from \eqref{FSconstraints}.
Without assuming the equations of motion, the transformations
\eqref{FS-SUSY} form an open algebra.

The transformation law \eqref{FS-SUSY}
can be rewritten for the chiral superfields $Q_\pm$ as follows:
\be
 \d Q_\pm = \pm\frac{1}{2} \bar \cD^2 (\bar \rho_\rL \bar Q_\mp )~.
\label{deltaQabeian}
\ee
The chiral superfields $Q_\pm$ have the same $\hat \cZ$-charge, 
\be
\hat \cZ Q_\pm = - \hf Q_\pm~.
\ee

%%%%%%%%%%%%%%%%%%%%%%%%%%%%%%%%%%%%%%%%%%%%

\subsection{Reduction of right projective multiplets to ${\mathbb M}^{3|4}_X$}

For the right projective multiplet, the technical details of their reduction to ${\mathbb M}^{3|4}_X$ are similar to the left ones, although there are some differences. The key differences from the left case are the following: (i) the analyticity conditions \eqref{anaN2} are replaced with \eqref{ana-right2}; and (ii) the ${\mathbb M}^{3|8}_X$-isometry transformation (\ref{IsometryGeneral}) for the right projective multiplets $Q_\mathrm{R}^{(n)}$ reads
\begin{equation}
    \delta_\xi Q_\mathrm{R}^{(n)} = \left( \xi + \Lambda^{\bai\baj}\mathbf{R}_{\bai\baj}\right)Q_\mathrm{R}^{(n)}\,.
\label{delta-Q-right}
\end{equation}

Throughout this subsection, we adopt the notation $\zeta := \zeta_\mathrm{R}$.

\subsubsection{Polar weight-$n$ multiplets} 

For the right polar weight-$n$ multiplet, reduced to ${\mathbb M}^{3|4}_X$, the isometry transformations (\ref{delta-Q-right}) may be written in the following explicit form:
\bsubeq
\bea 
\label{eq_dUpR}
\d_\x \U_\rR^{[n]}|
&=& \Big{[} 
\t +\ri t\Big(\z\frac{\pa}{\pa \z}-\frac{n}{2}\Big)
-\z{\ve}^\a \cD_\a
+\frac{1}{\z}\bar{\ve}_\a \cDB^\a
-\ri\s\Big(\z\frac{\pa}{\pa \z}-\frac{n}{2}\Big)
\non\\
&&~
+\Big( \z \bar{\L}_\rR 
+\frac{1}{\z}{\L}_\rR \Big)\z\frac{\pa}{\pa \z}
-n\,\z\bar{\L}_\rR 
 \Big{]} \U_\rL^{[n]}| 
 ~,~~~~~~
 \\  
\d_\x \breve{\U}_\rR^{[n]} |
&=& \Big{[} 
\t
+\ri t\Big(
\z\frac{\pa}{\pa \z}
+\frac{n}{2}\Big)
-\z{\ve}^\a \cD_\a
+\frac{1}{\z}\bar{\ve}_\a \cDB^\a
-\ri \s\Big(
\z\frac{\pa}{\pa \z}
+\frac{n}{2}\Big)
\non\\
&&~
+\Big(\z  \bar{\L}_\rR 
+\frac{1}{\z}{\L}_\rR \Big)\z\frac{\pa}{\pa \z}
+n \frac{1}{\z}{\L}_\rR 
 \Big{]}
 \breve{\U}_\rR^{[n]}  |
 ~.
\eea 
\esubeq
These transformations represent right polar multiplet counterparts of the the left transformation laws \eqref{eq_dUpL-aa}.

For comparison with the left case, it is instructive to consider the polar weight-one hypermultiplet
\begin{subequations}
\begin{align}
\U^{[1]}_\rR (\z)| &= \sum_{n=0}^\infty \z^n \U_{n} = \F + \z \Sigma + \cdots~, 
\qquad \bar \cD_\a \F =0~, \quad \bar \cD^2 \S =0~, \label{Taylor-r}
\\
\breve{\U}^{[1]}_\rR (\z)| &= \sum_{n=0}^\infty (-\z)^{-n} \bar \U_{n} ~.
\end{align}
\end{subequations}
The component superfields $\F:=\U_{0}$ and $\S := \U_{1}$
are chiral and complex linear, respectively. They describe physical $\cN=2$ multiplets. 
The other Taylor coefficients in \eqref{Taylor-r},  $\U_2, \U_3, \dots$, are complex unconstrained $\cN=2$ superfields. They describe auxiliary $\cN=2$ multiplets. 
The physical superfields have the following $\hat \cZ $ charges: 
\bea
\hat \cZ \F =  \hf \F~, \qquad \hat \cZ  \S = - \hf \S~.
\label{Zcharge-r}
\eea
By using the
transformation law (\ref{eq_dUpR})
one obtains the hidden off-shell supersymmetry transformations
\begin{subequations}
\begin{align}
\delta \F &= \bar{\ve}_\alpha \bar \cD^\alpha \Sigma
+ \L_\rR  \Sigma
	= - \frac{1}{2} \bar \cD^2 (\bar \rho_\rR  \Sigma)~, \\
\delta \Sigma &= -(\ve^\alpha \cD_\alpha 
+\bar \L_\rR) \F
	- \bar \cD_\alpha (\nrho^\alpha \U_{ 2}) ~.
	\end{align}
\end{subequations}
It is an instructive exercise to check that $\d \S$ is complex linear, $\bar \cD^2 \d \S=0$. 

\subsubsection{Real weight-$(2n)$ multiplets}
\label{SectO2-right}

A real right weight-$(2n)$ multiplet is given by its $\zeta$-decomposition analogous to the one in Eq.~(\ref{RealWeight-2n}),
\begin{subequations} \label{RealRightWeight-2n}
\bea
G_\rR^{(2n)}(v_\rR)&=& (\ri \, v^{\bar 1} v^{\bar 2})^n G_\rR^{[2n]}(\z)=( v^{\bar 1})^{2n} (\ri \,\z)^n G_\rR^{[2n]}(\z)~,\\
G_\rR^{[2n]}(\z) &=& 
\sum_{k=-p}^{p} G_k  \z^k~,
\qquad  {\bar G}_k = (-1)^k G_{-k} ~.
\eea
\end{subequations}
For this multiplet, the isometry transformation (\ref{delta-Q-right}) takes the following form:
\bea
\d_\x G_\rR^{[2n]} |
&=&\Big{[}
\t
+\ri t\z\frac{\pa}{\pa \z}
-\z{\ve}^\a \cD_\a
+\frac{1}{\z}\bar \ve_\a \cDB^\a
-\ri \s\z\frac{\pa}{\pa \z}
\non\\
&&~
+\Big(  \bar{\L}_\rR\z+ \L_\rR\frac{1}{\z} \Big)\z\frac{\pa}{\pa \z}
+ n\Big( \L_\rR\frac{1}{\z}- \bar{\L}_\rR\z \Big)
 \Big{]} G_\rR^{[2n]} |
 ~.
 \label{L2n-right}
\eea 

As an example, consider a right ${\cal O}(2)$ multiplet represented by a real superfield $W^{\bai\baj} = W^{\baj\bai} = \overline{W_{\bai\baj}}$ obeying the constraints
\begin{equation}
    {\cal D}_\alpha^{i(\bai} W^{\bak\bal)} =0\,.
\end{equation}
The corresponding real projective right weight-2 superfield $W_\mathrm{R}^{(2)}(v) = W^{\bai\baj} v_\bai v_\baj$ has the following $\zeta$-decomposition
\begin{subequations}
\bea
W^{(2)}_\rR(v)
&=&
\ri (v^{\bar 1})^2 \z W^{[2]}_\mathrm{R}(\z)~, \\
W^{[2]}_\mathrm{R}(\z)|
&=&
\frac{1}{\z}\varphi + G
-\z\bar\varphi
~, \qquad \bar \cD_\a \varphi = 0 ~, \qquad \bar \cD^2 G=0~,
\eea
\end{subequations}
where the chiral superfield $\varphi$ and the real linear superfield $G$ are associated with the following components of $W^{\bar i\bar j}$:
\be
\varphi:=-\ri \, W^{\bar 2\bar 2}|~,\qquad
G:=2\ri \,W^{\bar 1\bar 2}|~,\qquad
\bar\varphi=\ri \,W^{\bar 1\bar 1}|
~.
\ee
The transformation law (\ref{L2n-right}) dictates the $\hat{\cal Z}$-charges of these superfields,
\be 
\hat \cZ \varphi = \varphi ~, \qquad \hat \cZ G = 0~,
\ee
and their non-manifest supersymmetry transformations:
\begin{equation}
    \delta\varphi = -\frac12\bar{\cal D}^2(\bar\rho_\mathrm{R} G)\,,\qquad
    \delta G = -{\cal D}^\alpha(\varepsilon_\alpha \varphi) - \bar{\cal D}_\alpha (\bar\varepsilon^\alpha \bar\varphi)\,.
\end{equation}
Using the identities (\ref{id4.17})--(\ref{id4.22}) these variations may be expressed via the parameters $\rho_\mathrm{L}$ and $\bar\rho_\mathrm{L}$,
\begin{equation}
    \delta\varphi = \frac\ri{8X}\bar{\cal D}^2(G{\cal D}^2\rho_\mathrm{L})\,,\qquad
    \delta G = {\cal D}^\alpha \bar{\cal D}_\alpha(\rho_\mathrm{L} \bar\varphi - \bar\rho_\mathrm{L} \varphi)\,.
\label{HiddenSusyAbelian}
\end{equation}
This form of non-manifest supersymmetry transformation appears useful because the superfield $W^{\bar i\bar j}$ plays the role of the field strength of the \emph{left} $\cN=4$ vector multiplet, see Sec.~\ref{Sec6.2} below.

\subsubsection{On-shell Fayet-Sohnius hypermultiplet} 

The on-shell right Fayet-Sohnius hypermultiplet is described by a right isospinor superfield $q^{\bar i} (z) $ and its conjugate  $\bar q_{\bar i} (z)$. The former is constrained by
\be
\cD_\a^{i (\bar i} q^{\bar j ) } =0~.
\ee
Associated with $q^{\bar i}$ are $\cN=2$ chiral $Q_+$ and antichiral $\bar Q_-$ superfields 
\be
Q_+ = q^{\bar 2}|~, \qquad \bar \cD_\a Q_+ = 0~; \qquad 
\bar Q_- = q^{\bar 1}|~, \qquad \cD_\a \bar Q_- =0
~,
\ee
with the following  $\hat \cZ$ charges 
\be
\hat \cZ Q_\pm =  \hf Q_\pm~, \qquad \hat \cZ \bar Q_\pm = -\hf \bar Q_\pm~.
\ee
The hidden supersymmetry transformation acts on them by the rule
\be
\d Q_\pm = \pm \frac{1}{2} \bar \cD^2 (\bar \rho_\rR \bar Q_\mp )~, \qquad 
\d \bar Q_\pm = \pm\frac{1}{2} \cD^2 ( \rho_\rR Q_\mp)~.
\ee

%%%%%%%%%%%%%%%%%%%%%%%%%%%%%%%%

\section{Theories with non-centrally extended  $\cN=4$ Poincar\'e supersymmetry}

In order to formulate the dynamics of rigid $\cN=4$ supersymmetric field theories in 
 ${\mathbb M}^{3|8}_X$,  a manifestly supersymmetric action principle is required. 
Such an action principle is obtained 
by restricting to  ${\mathbb M}^{3|8}_X$ the locally supersymmetric action introduced in \cite{KLT-M11}.
For a large family of theories without higher derivatives the action looks like
\bea
S= S_{\rm left} + S_{\rm right} ~, 
\eea
where $S_{\rm left} $ ($S_{\rm right} $) is constructed in terms left (right) projective multiplets.\footnote{More general (hybrid) $\cN=4$ supersymmetric actions exist that involve both left and right projective multiplets \cite{KLT-M11}. Typically they describe higher-derivative terms.}

The left action is generated by a real Lagrangian  $\cL_\rL^{(2)} (z,v_\rL)$, 
which is a real left projective  weight-2  multiplet. The action can be written in a manifestly $\cN=4$ supersymmetric form on ${\mathbb M}^{3|8}_X$, see \cite{KLT-M11,BKTM} for the technical details, or it may be reduced to the deformed $\cN=2$ Minkowski superspace  ${\mathbb M}^{3|4}_X$, in complete analogy with the AdS analysis of \cite{BKTM}. In this paper we will need only the latter representation for the action 
\bea
S(\cL_\rL^{(2)}) &=& \int \rd^{3|4}z
\, 
\oint_C \frac{\rd \zeta}{2\pi \ri \zeta}\, \cL^{[2]}_\rL~, \qquad
\cL_\rL^{(2n)}(v_\rL)=\ri \,\z ( v^1)^{2}  \cL_\rL^{[2n]}(\z_\rL)~,
\label{LeftAction}
\eea
where we have denoted $\rd^{3|4} z= \rd^3x\, \rd^2{\q} \rd^2{\qb}$.
The action may be shown to be invariant under the $\cN=4 $ isometry transformations \eqref{harmult1} which act on 
$ \cL^{[2]}_\rL$ by the rule 
\bea
\d_\x \cL^{[2]}_\rL 
&=&\Big{[}
\t
+\ri t\z\frac{\pa}{\pa \z}
+\z\bar{\ve}^\a \cD_\a
-\frac{1}{\z}\ve_\a \cDB^\a
+\ri \s\z\frac{\pa}{\pa \z}
\non\\
&&~
+\Big(  \bar{\L}_\rL\z+ \L_\rL\frac{1}{\z} \Big)\z\frac{\pa}{\pa \z}
+ \Big( \L_\rL\frac{1}{\z}- \bar{\L}_\rL\z \Big)
 \Big{]} \cL_\rL^{[2]} 
 ~,
\eea 
see \cite{BKTM} for the proof in AdS.

%%%%%%%%%%%%%%%%%%%%%%%%%%%%%%%%%%%%%%

\subsection{Supersymmetric nonlinear sigma models}

General nonlinear sigma models  with non-centrally extended  $\cN=4$ Poincar\'e supersymmetry are deformations of the $\cN=4$ superconformal sigma models \cite{BKTM}. 

\subsubsection{Off-shell approach}

Let us briefly elaborate on the left $\s$-model sector.  
The dynamical variables are weight-one arctic multiplets $\U_\rL^{(1) I} (v_\rL) $
and their smile-conjugate antarctic multiplets $\breve \U_\rL^{(1) \bar I}(v_\rL)$.
The $\s$-model Lagrangian $\cL^{(2)}_\rL$ defining the $\cN=4$ supersymmetric 
action  \eqref{LeftAction} is 
\bea
\cL^{(2)}_\rL = \ri {\mathfrak K}_\rL (\U^{(1)}_\rL , \breve \U^{(1)}_\rL) ~\implies ~
 \cL^{[2]}_\rL (\U_\rL, \breve{\U}_\rL;\z) = \frac{1}{\z} 
 {\mathfrak K}_\rL (\U_\rL , \z\breve \U_\rL) ~,
\eea
where we have denoted  $\U_\rL^{ [1] } (\z) $ and $\breve \U_\rL^{ [1] }(\z)$ 
 simply as $\U_\rL(\z) $ and  $\breve \U_\rL(\z)$, respectively. 
Here 
${\frak K} (\F , \bar \O )$ is a homogeneous function of $2n$ 
complex variables $\F^I $ and $\bar \O^{\bar J} $, 
\bea
\Big( \F^I \frac{\pa}{\pa \F^I} + \bar  \O^{\bar I} \frac{\pa}{\pa \bar \O^{\bar I} } \Big) {\frak K} (\F, \bar \O )
= 2 {\frak K} (\F , \bar \O )~,
\label{hom_cond_strange}
\eea
subject to the reality condition
\bea
\bar {\frak K} (\bar \F , -\O ) = -{\frak K} (\O , \bar \F )~,
\label{reality_condition_strange}
\eea
where $\bar {\frak K} (\bar \F ,  \O)$ denotes the complex conjugate of  ${\frak K} ( \F , \bar \O )$. This reality condition is required in order for $\cL^{(2)}_\rL$ to be real under the smile conjugation. 

The arctic  $\U_\rL^I (\z) $ and antarctic $\breve\U_\rL^{\bar I} (\z)$ multiplets
are given by  series in $\z$ 
\begin{align}
\U_\rL^I(\z) = \sum_{n=0}^\infty \z^n \U_{\rL n}^I = \Phi_\rL^I + \z \Sigma_\rL^I + \cdots~, \qquad
\breve{\U}_\rL^{\bar I}(\z) = \sum_{n=0}^\infty (-\z)^{-n} \bar \U_{\rL n}^{\bar I} ~.
\end{align}
The component superfields $\Phi_\rL^I:=\U_{\rL 0}^I$ and $\S_\rL^I := \U_{\rL 1}^I$
are chiral and complex linear respectively. These correspond to the
physical fields while the remaining superfields are auxiliary.
This off-shell formulation possesses an infinite number of auxiliary fields.
 The auxiliary superfields, 
$ \U_{\rL 2}^I,  \U_{\rL 3}^I, \dots$, can be expressed  in terms of the physical ones, 
using their equations of motion 
\bea
\frac{\pa {\mathbb L}_\rL}{\pa \U^I_{\rL n}} = 0 ~, \quad n = 2,3, \dots~, \qquad 
\mathbb L_\rL := \oint_C \frac{\rd\zeta}{2\pi \ri \z} \cL^{[2]}_\rL~.
\label{EoM}
\eea
The price for such a simplification 
is that $\cN=4$ supersymmetry is no longer off-shell.
This leaves the Lagrangian 
$\mathbb L_\rL= {\mathbb L}_\rL(\F_\rL, \S_\rL, \bar \F_\rL, \bar \S_\rL )
$ depending 
only on $\Phi_\rL$, $\Sigma_\rL$ and
their complex conjugates. 
This Lagrangian is invariant under the central charge transformation \eqref{Zcharge}, which means 
\bea
\Big( \Phi_\rL^I \frac{\pa}{\pa \Phi_\rL^I}
- \S_\rL^I \frac{\pa}{\pa \S_\rL^I} \Big)
\mathbb L_\rL
= \Big(\bar\F_\rL^{\bar I} \frac{\pa  }{\pa \bar\F_\rL^{\bar I}} -
\bar\Sigma_\rL^{\bar I} \frac{\pa  }{\pa \bar\Sigma_\rL^{\bar I}} \Big)\mathbb L_\rL~.
\label{CC}
\eea
The auxiliary superfields $ \U_{\rL 2}^I,  \U_{\rL 3}^I, \dots$, and their conjugates do not contribute because they satisfy the algebraic equations of motion \eqref{EoM}.
On the same ground, the homogeneity condition \eqref{hom_cond_strange} is equivalent to 
\bea
\Big(  \Phi^I \frac{\pa }{\pa \Phi^I}+ \Sigma^I \frac{\pa }{\pa \Sigma^I}
	+ \bar\Phi^{\bar I} \frac{\pa }{\pa \bar\Phi^{\bar I}}
	+ \bar\Sigma^{\bar I} \frac{\pa }{\pa \bar\Sigma^{\bar I}} \Big) {\mathbb L}_\rL =2 {\mathbb L}_\rL ~.
\eea
Combining the last two relations leads to
\begin{align}
\Big( \Phi_\rL^I \frac{\pa}{\pa \Phi_\rL^I}
	+ \bar\Sigma_\rL^{\bar I} \frac{\pa  }{\pa \bar\Sigma_\rL^{\bar I}} \Big)  \mathbb L_\rL
	=\mathbb L_\rL 	~.
	\label{HomCond}
\end{align}

The complex linear superfields $\Sigma_{\rL}^{ I}$ and their conjugates
$\bar \Sigma_{\rL}^{\bar I}$  can be dualised into chiral superfields $\Psi_{\rL I}$ 
and their conjugates $\bar \J_{\rL \bar I}$  via a Legendre transformation. Then we arrive at the dual action
\begin{align}
S_{\rm dual} = \int \rd^{3|4}z \, \mathbb K_\rL(\Phi_\rL,  \Psi_\rL, \bar\Phi_\rL,\bar\Psi_\rL)
~, \qquad	\mathbb K_\rL = \mathbb L_\rL + \S_\rL^I \Psi_{\rL I} 
+ \bar\S_\rL^{\bar J} \bar\Psi_{\rL \bar J}~.
\label{10.13}
\end{align}
The target space of this $\s$-model is a hyperk\"ahler cone. 
The cone condition follows from the fact that the K\"ahler potential $\mathbb K_\rL$ 
obeys the homogeneity condition
\begin{align}\label{eq_Kcone}
 \phi_\rL^\ra \frac{\pa}{\pa \phi_\rL^\ra}   \mathbb K_\rL = \mathbb K_\rL ~, 
\qquad  \phi_\rL^\ra = (\Phi_\rL^I, \Psi_{\rL I})~,
\end{align}
as a consequence of \eqref{HomCond}.
In accordance with \eqref{Zcharge}, the chiral superfields $\phi_\rL^\ra$ carry the same $\hat \cZ$-charge,
\bea
\hat \cZ \phi_\rL^\ra = -\hf \phi_\rL^\ra~.
\eea
The target space is hyperk\"ahler since it possesses a 
covariantly constant holomorphic two-form $\omega_{\ra\rb}$  given by
\begin{align}\label{eq_canonicalOmega}
\omega_{\ra\rb} =
\begin{pmatrix}
0 & \delta_I{}^J \\
-\delta^I{}_J & 0
\end{pmatrix}
\end{align}
with the property $\omega^{\ra\rb} \omega_{\rb\rc} = -\delta^\ra_\rc$, 
where $ \omega^{\ra \rb} = g^{\ra \bar \rc } g^{\rb \bar \rd} \bar \o_{\bar \rc \bar \rd}$
and $g^{\ra \bar \rb }$ is the inverse of the K\"ahler metric. 
The action \eqref{10.13} proves to be invariant under the extended supersymmetry
transformation
\begin{align}
\delta \Phi_\rL^I
	= - \frac{1}{2} \bar \cD^2 \Big(\bar \rho_\rL \frac{\pa\mathbb K_\rL}{\pa \Psi_{\rL I}}\Big)~, \qquad
\delta \Psi_{\rL I}
	= \frac{1}{2} \bar \cD^2 \Big(\bar \rho_\rL \frac{\pa\mathbb K_\rL}{\pa \Phi_{\rL}^I}\Big)~,
\end{align}
which we can cast in the target-space reparametrization-covariant form \cite{BKTM}
\begin{align}
\delta \phi_\rL^\ra = - \frac{1}{2} \bar \cD^2 \Big(\bar \rho_\rL \omega^{\ra\rb} 
\pa_\rb \mathbb K_{\rL}\Big)~.
\end{align}

It should be remarked that the
reality condition \eqref{reality_condition_strange}
has the simplest form in the case 
that ${\frak K} (\F , \bar \O ) = K (\F , \bar \O )$, where $K (\F , \bar \O )$ is 
homogeneous with respect to $\F$ (or, equivalently, 
with respect to  $\bar \O$).
Then the reality condition \eqref{reality_condition_strange} is equivalent to 
$\bar {K} (\bar \F , \F ) = {K} (\F , \bar \F )$, that is ${K} (\F , \bar \F )$
is real. The homegeneity conditions obeyed by
$K(\F, \bar \F)$ are:
\bea
 \F^I \frac{\pa}{\pa \F^I} 
K(\F, \bar \F) =  K( \F,   \bar \F) ~, \qquad
 \bar \F^{\bar I} \frac{\pa}{\pa \bar \F^{\bar I}} 
K(\F, \bar \F) =  K( \F,   \bar \F) ~.
\label{Kkahler}
\eea
The supersymmetric nonlinear $\s$-model \eqref{10.13} generated by 
the Lagrangian
\bea
\cL^{[2]}_\rL (\U_\rL, \breve{\U}_\rL;\z) =
\frac{1}{\z}  {\mathfrak K}_\rL (\U_\rL , \z\breve \U_\rL) = K ( \U_\rL , \breve \U_\rL)~,
\eea
has a simple geometric interpretation. 
This theory is associated with a K\"ahler cone $\cX$ for which $K(\F, \bar \F)$ is the 
preferred K\"ahler potential, see Appendix \ref{Cone}. The $\s$-model target space turns out to be (an open domain of)
the cotangent bundle $T^*\cX$ which is a hyperk\"ahler cone \cite{K-duality}.
In the 4D $\cN=2$ case, 
off-shell  supersymmetric nonlinear $\s$-models on the cotangent bundles of K\"ahler manifols were studied in 
\cite{GK1,GK2,AN,AKL1,AKL2,KN}. 

%%%%%%%%%%%%%%%%%%%%%%%%%%%%%%%%%%%%

\subsubsection{On-shell approach}

We now describe the most general nonlinear $\s$-model with non-centrally extended $\cN=4$ Poincar\'e supersymmetry by using an on-shell approach. 
First, we give its realisation  in terms of chiral superfields on ${\mathbb M}^{3|4}_X$, and then we present its component form. 

The  target space of the $\s$-model is a hyperk\"ahler cone $\cM= \cM_{\rL} \times \cM_{\rR}$, see Appendix \ref{Cone} 
for a brief discussion of hyperk\"ahler cones. 
This direct product structure is manifested by the presence of two copies of the covariantly constant
holomorphic two-form $\omega_{\ra\rb}$, which we denote
$\omega_{\rL\ra\rb}$ and $\omega_{\rR\ra\rb}$. They obey both an orthogonality condition
and a completeness condition
\begin{align}\label{omegas}
\omega_\rL^{\ra\rb} \omega_{\rR\rb\rc} = 0~, \qquad
\omega_\rL^{\ra\rb} \omega_{\rL\rb\rc} + \omega_\rR^{\ra\rb} \omega_{\rR\rb\rc} = -\delta^\ra{}_\rc~.
\end{align}
These conditions allow us to construct covariantly constant projection operators
\begin{gather}
(P_\rL)^\ra{}_\rb = - \omega_\rL^{\ra\rc} \omega_{\rL \rc\rb}~, \qquad
(P_\rR)^\ra{}_\rb = - \omega_\rR^{\ra\rc} \omega_{\rR \rc\rb}~, \eol
P_\rL P_\rR = 0~, \qquad P_\rL + P_\rR = {\mathbbm 1}~.\label{projs}
\end{gather}
The $\omega_\rL$ and $\omega_\rR$ arise from the covariantly constant holomorphic two-form
$\omega_{\ra\rb} := \omega_{\rL\ra\rb} + \omega_{\rR\ra\rb}$ which can be
easily seen to obey $\omega^{\ra\rb} \omega_{\rb\rc} = -\delta^\ra_\rc$.
It holds that  $\omega_{\rL\ra\rb} = (P_\rL)_\ra{}^\rc \omega_{\rc\rb}$ and similarly
for $\omega_{\rR\ra\rb}$.

Let $\c$ be  the homothetic conformal Killing vector on the hyperk\"ahler cone $\cM$,  
 \bea 
\c = \c^\ra (\f) \frac{\pa}{\pa \f^\ra} + {\bar \c}^{\bar \ra} (\bar \f)  \frac{\pa}{\pa {\bar \f}^{\bar \ra}}~.
\eea
Its holomorphic component  $\c^\ra$ can be decomposed into
left and right sectors via
\begin{align}
\chi_\rL^\ra = (P_\rL)^\ra{}_\rb \chi^\rb~, \qquad
\chi_\rR^\ra = (P_\rR)^\ra{}_\rb \chi^\rb~,
\end{align}
so that the K\"ahler potential is given by
\begin{align}
K = \chi^\ra \chi_\ra = \chi_\rL^\ra \chi_{\rL \ra} + \chi_\rR^\ra \chi_{\rR \ra} =K_\rL + K_\rR~.
\end{align}
Here $K_\rL =  \chi_\rL^\ra \chi_{\rL \ra} $ is the hyperk\"ahler potential for the
left sector as is $K_\rR =\chi_\rR^\ra \chi_{\rR \ra}$ for the right sector.

To comply with  the notation of \cite{BKTM}, we identify the central charge operator 
with $\Delta := \ri (\bL^{1 2} - \bR^{\bar 1 \bar 2}) = \ri \hat \cZ$. The vector field 
\bea
\D =  \D^\ra (\f) \frac{\pa}{\pa \f^\ra} + {\bar \D}^{\bar \ra} (\bar \f)  \frac{\pa}{\pa {\bar \f}^{\bar \ra}}= \ri \hat \cZ
\eea
must annihilate the K\"ahler potential, 
\begin{align}
\D K=\Delta^\ra K_\ra + \bar \Delta^{\bar \ra} K_{\bar \ra} = 0~.
\end{align}
It turns out that 
$\Delta^\ra$ is given by
\begin{align}
\Delta^\ra = -\frac{\ri}{2} \chi_\rL^\ra + \frac{\ri}{2} \chi_\rR^\ra~, 
\end{align}
since the non-manifest supersymmetry transformation is
\begin{align}\label{susy}
\delta \phi^\ra = -\frac{1}{2} \bar \cD^2 (\bar \rho_\rL \Omega_\rL^\ra)
	-\frac{1}{2} \bar \cD^2 (\bar \rho_\rR \Omega_\rR^\ra)~,
\end{align}
where the functions $\Omega_\rL^\ra (\f , \bar \f) $ and $\Omega_\rR^\ra (\f , \bar \f) $ are such that $\omega_{\rL \ra\rb} = g_{\ra \bar \ra} \pa_\rb \Omega_\rL^{\bar \ra}$
and $\omega_{\rR \ra\rb} = g_{\ra \bar \ra} \pa_\rb \Omega_\rR^{\bar \ra}$. 
No superpotential $W(\f)$ is allowed by the non-centrally extended $\cN=4$ Poincar\'e supersymmetry.\footnote{This is true for a generic sigma model in ${\rm AdS}^X_{(3|4,0)}$ with $|X|\ne 2S$. Remarkably, in the critical case with $X=\pm2S$, where either the $\sSU(2)_{\rm L}$ or $\sSU(2)_{\rm R}$ symmetry disappears from the structure group of ${\rm AdS}^X_{(3|4,0)}$, a superpotential can be introduced whenever the target space possesses a tri-holomorphic isometry \cite{BKTM}.} Thus the $\s$-model superfield Lagrangian is 
\bea
K(\f , \bar \f) = K_\rL(\f, \bar \f) + K_\rR(\f, \bar \f)~.
\eea

The component form of the supersymmetric nonlinear $\s$-model may be easily derived.  
The components of $\phi^\ra$ are defined by 
\bsubeq
\begin{align}
\varphi^\ra &:= \phi^\ra\vert~, \qquad
\psi_\alpha^\ra := \frac{1}{\sqrt 2} \cD_\alpha \phi^\ra \vert~, \\
F^\ra &:= -\frac{1}{4} g^{\ra \bar \rb} \cD^2 K_{\bar \rb} \vert
	= -\frac{1}{4} (\cD^2 \phi^\ra + \Gamma^\ra{}_{\rb\rc} \cD^\alpha \phi^\rb \cD_\alpha \phi^\rc)\vert
~.
\end{align}
\esubeq
In particular, the auxiliary field $F^\ra$ transforms covariantly under 
target space reparametrisations.
The vector derivative on the fermion is similarly reparametrisation covariant,
\begin{align}
\hat \cD_m \psi_\alpha^\ra := \cD_m \psi_\alpha^\ra 
+ \Gamma^\ra{}_{\rb\rc} \cD_m \varphi^\rb \psi_\alpha^\rc~,
\end{align}
and the action of $\D$ on the physical fields is defined as  
\bea
\D \vf^\ra = \D^\ra(\vf), \qquad \D \j^\ra_\a = \j^\rb_\a \pa_\rb \D^\ra (\vf)~.
\eea 

The component Lagrangian is
\begin{align}
L &= -g_{\ra \bar \ra} \cD_m \varphi^\ra \cD^m \bar\varphi^{\bar \ra}
	- \ri g_{\ra \bar \ra} \bar\psi_\alpha^{\bar \ra} \hat \cD^{\alpha \beta} \psi_\beta^{\ra}
	+ \frac{1}{4} R_{\ra \bar \ra \rb \bar \rb} (\psi^{\ra} \psi^\rb) (\bar\psi^{\bar \ra} \bar\psi^{\bar \rb})
	\eol & \quad
	- \frac{\ri}{2} X (\psi^\ra \bar\psi^{\bar \rb}) (P_\rL - P_\rR)_{\ra \bar \rb}
	- \frac{1}{4} X^2 (K_\rL + K_\rR)~.
\label{12.8}
\end{align}
We emphasize that the action
is \emph{not} superconformal even though the target space is a cone.

The Lagrangian \eqref{12.8} describes the most general nonlinear $\s$-model with
the non-centrally extended  $\cN=4$ Poincar\'e supersymmetry. 
Setting $X=0$ in  \eqref{12.8}
gives the most general $\cN=4$ superconformal $\s$-model, 
with its target space being a hyperk\"ahler cone $\cM_{\rL} \times \cM_{\rR}$.
The deformation parameter $X$ appears in both terms in the second line of \eqref{12.8}. 
The first structure constitutes the fermionic mass term, while the second gives  
the scalar potential 
\bea
V = \frac{1}{4} X^2 (K_\rL + K_\rR)  ~.
\label{12.9}
\eea
For each of the left and right $\s$-model sectors, 
the scalar potential is constructed in terms of the 
homothetic  conformal Killing vector associated with the corresponding target space. 
This follows from the general result that for any hyperk\"ahler cone 
the preferred K\"ahler potential is given in terms of the homothetic 
conformal Killing vector $\c$ as follows:
\bea
K (\f, \bar \f) :={ g}_{\ra \bar \rb}(\f, \bar \f)  \, \c^\ra (\f) {\bar \c}^{\bar \rb} (\bar \f)~.
\eea

%%%%%%%%%%%%%%%%%%%%%%%%%%%%%%%%%%%%%%%

\subsection{Vector multiplet models} 
\label{Sec6.2}

There are two types of $\cN=4$ vector multiplets in three dimensions, left and right ones. 
An Abelian left vector multiplet is described by an unconstrained gauge prepotential
that is  a left  tropical weight-zero multiplet $V_\rL(v_\rL)$, which is defined modulo gauge transformations 
\bea
\d V_\rL = \ri ( \breve{\l}_\rL -\l_\rL) ~, 
\label{LeftGaugeTr}
\eea
where the gauge parameter $\l_\rL (v_\rL) $ is a left arctic weight-zero multiplet.
Associated with $V_\rL$ is a gauge-invariant field strength \cite{KLT-M11, KTM14} defined by
\bea \label{LeftFieldStrength}
W^{\bai \baj}
&=&
\frac{\ri}{4}
\cD^{ij \bai  \baj}
\oint_\g \frac{(v_\rL, \rd v_\rL)}{2\pi}
\frac{u_i u_j}{(v_\rL ,u_\rL)^2}
V_\rL(v_\rL) 
=W^{\baj\bai} = \overline{W_{\bai\baj}}~,
\eea
where we have introduced the operator
\bea
&&\cD^{ij \bai  \baj}:=\cD^{\a (i ( \bai}\cD_\a^{j)\baj)}
~.~~~~~~
\eea
This  field strength is a right linear multiplet, 
\bea
\cD_\a^{i(\bai}W^{\bak\bal)}=0~.
\eea
Making use of $W^{\bai \baj}$ allows us to introduce the real right $\cO(2)$ multiplet 
$W_\rR^{(2)}:=v_\bai v_\baj W^{\bai\baj}$.

An Abelian right vector multiplet is defined similarly. It  is described by a right  tropical weight-zero multiplet $V_\rR(v_\rR)$, which is defined modulo gauge transformations 
\bea
\d V_\rR = \ri ( \breve{\l}_\rR -\l_\rR) ~, 
\eea
with the gauge parameter $\l_\rR (v_\rR) $ being a right arctic weight-zero multiplet.
Associated with $V_\rR$ is the gauge-invariant field strength 
\bea
W^{i j}
&=&
\frac{\ri}{4}
\cD^{ij \bai  \baj}
\oint_\g \frac{(v_\rR, \rd v_\rR)}{2\pi}
\frac{u_\bai u_\baj}{(v_\rR ,u_\rR)^2}
V_\rR(v_\rR) 
=W^{j i} = \overline{W_{ij}}~,
\eea
which is a left linear multiplet, 
\bea
\cD_\a^{(i\bai}W^{kl)}=0~.
\eea
The index-free superfield $W_\rL^{(2)}:=v_iv_jW^{ij}$ is a real left $\cO(2)$ multiplet. 

There are two large families of interacting supersymmetric field theories in ${\mathbb M}^{3|8}_X$. 
One of them consists of deformations of $\cN=4$ superconformal field theories in  ${\mathbb M}^{3|8}$, while the other includes deformations of those $\cN=4$ supersymmetric gauge theories in ${\mathbb M}^{3|8}$
which are not superconformal but possess the $R$-symmetry group $\sSU(2)_\rL \times \sSU(2)_\rR$.  The vector-multiplet theories in the second family  are gauge invariant and massive, and therefore they should contain Chern-Simons terms at the component level.  

It is worth giving examples of models that belong to the first family. Let $W^{(2)}_I $ be the left  $\cO(2)$  field strengths for $n$ right Abelian vector multiplets, with $I=1, \dots , n$.
A gauge-invariant action functional is generated by a Lagrangian of the form
\bea
\cL^{(2)}_\rL &=& 
\cF_\rL (W^{(2)}_I ) ~, \qquad W^{(2)}_I \frac{\pa }{\pa W^{(2)}_I} \cF_\rL =\cF_\rL~.
\label{VMmodel}
\eea
In the $n=1$ case, there is a unique choice for $\cL^{(2)}_\rL$:
\bea
\cL^{(2)}_\rL = -W^{(2)}_\rL \ln \frac{W^{(2)}_\rL }{\ri {\bm \Upsilon}^{(1)}_\rL \breve{\bm \Upsilon}^{(1)}_\rL } ~,
\label{VMmodel-improved}
\eea
where it is assumed that $W_\rL:= \sqrt{ W^{ij} W_{ij}} \neq 0$.
The arctic multiplet ${\bm \Upsilon}^{(1)}_\rL $ may be shown to be a purely gauge degree of freedom.\footnote{The vector-multiplet Lagrangian \eqref{VMmodel} is a 3D analogue of the model given in \cite{K-08} to describe the 4D $\cN=2$ improved tensor multiplet \cite{deWPV} in curved projective superspace.} 
Another example is provided by the following model (with $e$ the hypermultiplet charge)
\bea
\cL^{(2)}_\rL &=& \ri 
\breve{ \Upsilon}^{(1)}_\rL {\rm e}^{e V_\rL} { \Upsilon}^{(1)}_\rL
-W^{(2)}_\rL \ln \frac{W^{(2)}_\rL }{\ri {\bm \Upsilon}^{(1)}_\rL \breve{\bm \Upsilon}^{(1)}_\rL } ~,
\eea
which is a deformation of $\cN=4$ superconformal electrodynamics. 
This Lagrangian is invariant under the gauge transformation \eqref{LeftGaugeTr} accompanied by the variation
\bea
\d { \Upsilon}^{(1)}_\rL = e \l_\rL { \Upsilon}^{(1)}_\rL~.
\eea

Computing the contour integral 
\bea
\mathbb L_\rL = \oint_C \frac{\rd\zeta}{2\pi \ri \z} \cL^{[2]}_\rL~.
\eea
for the model \eqref{VMmodel} leads to a Lagrangian
${\mathbb L}_\rL \equiv L(G_I, \vf_I , \bar \vf_I) $ satisfying the equations
\begin{subequations}
\bea
\Big( G_I \frac{\pa }{\pa G_I} + \vf_I \frac{\pa }{\pa \vf_I} + \bar \vf_I \frac{\pa }{\pa \bar \vf_I}
\Big) L &=& L  ~, \\
\vf_I \frac{\pa L}{\pa \vf_I} - \bar \vf_I \frac{\pa L}{\pa \bar \vf_I} &=&0~,\\
\frac{\pa^2 L}{\pa G_I  \pa  \vf_J} - \frac{\pa^2 L}{\pa G_J  \pa  \vf_I} &=&0~.
\eea
\end{subequations}
For the model \eqref{VMmodel-improved} we obtain
\begin{align}
L(G, \vf, \bar \vf) =  \sqrt{ G^2 +4\vf \bar \vf }
    - G \, \ln \left(\frac{G +   \sqrt{ G^2 +4\vf \bar \vf }}{\sqrt{4 \vf \bar\vf}}\right)
    ~.
    \label{LImproved}
\end{align}
This model is a 3D analogue of the improved $\cN=2$ tensor multiplet studied in, e.g.,  
\cite{LR, HitchinKLR, deWRV,Butter:2011kf}. In Ref.~\cite{BPS2} it was shown that the Lagrangian (\ref{LImproved}) appears in the low-energy effective action of certain 3D Yang-Mills gauge theories with chiral matter in $\cN=2$ superspace.

Now we give examples of interacting supersymmetric field theories in ${\mathbb M}^{3|8}_X$ that are  deformations of those $\cN=4$ supersymmetric field theories in ${\mathbb M}^{3|8}$
which possess the $R$-symmetry group $\sSU(2)_\rL \times \sSU(2)_\rR$. Their purely vector multiplet sectors are Chern-Simons type models. Starting from the left vector multiplet, $V_\rL(v_\rL)$,  and the corresponding field strength \eqref{LeftFieldStrength}, one can introduce the descendant 
\bea
{\bm W}^{i j} :=-\frac{\ri}{12}\cD^{ij\bai\baj}W_{\bai \baj}~, 
\label{Compos}
\eea
which proves to be a left linear multiplet \cite{KTM14},
\bea
\cD_\a^{(i\bai} {\bm W}^{kl)}=0~.
\eea
The corresponding real left $\cO(2)$ multiplet, 
${\bm W}_\rL^{(2)}:=v_iv_j {\bm W}^{ij}$, can be used to construct
 $\cN=4$ supersymmetric electrodynamics  in  ${\mathbb M}^{3|8}_X$
\bea
\cL^{(2)}_\rL &=& \ri 
\breve{ \Upsilon}^{(1)}_\rL {\rm e}^{e V_\rL} { \Upsilon}^{(1)}_\rL
+\frac1{2g^2} V_\rL {\bm W}^{(2)}_\rL ~,
\label{super-ED}
\eea
where $g$ is a coupling constant of mass dimension 1/2. This model was earlier constructed for $(p,q)$ AdS superspaces with $p+q=4$ \cite{KTM14}.{\footnote{More precisely, the projective action principle representation described by the second term in \eqref{super-ED}, based on \eqref{Compos}, is a new result of our paper. In contrast, in \cite{KTM14}, an equivalent representation in terms of a full superspace integral was used.} 
The second term in \eqref{super-ED}
is modelled on the $\cN=4 $ BF term 
\cite{BrooksG}.
\bea
\cL^{(2)}_\rL = V_\rL {W}^{(2)}_\rL \quad \Longleftrightarrow \quad
\cL^{(2)}_\rR = V_\rR {W}^{(2)}_\rR~,
\eea
which involves two vector multiplets, left and right ones.

An important comment is in order. For any supergravity background, there is an alternative procedure \cite{KN14}, inspired by the 4D $\cN=2$ analysis in \cite{Butter:2010jm}, to construct a composite left linear multiplet, $\bm G{}^{{i} {j}} $, from the left vector multiplet:
\bea
\bm G{}^{{i}{j}} 
= \frac{\ri}{4} (\cD^{i j \bar{i} \bar j} 
+ 8 \ri \cS^{ij\bar{i}\bar{j}} ) \Big( \frac{W_{\bai \baj}}{W_\rR}\Big) ~, \qquad 
W_\rR:= \sqrt{ W^{\bai \baj} W_{\bai \baj}}~.
\label{G}
\eea
It is applicable only in the case when $W_\rR$ is nowhere vanishing, 
$W_\rR \neq 0$. The superfield $\bm G{}^{{i}{j}} $ proves to be primary 
under the super-Weyl transformations \cite{KN14}. Unlike $\bm G{}^{{i}{j}} $, 
our composite linear multiplet \eqref{Compos} exists only in the AdS superspaces and ${\mathbb M}^{3|8}_X$. 
Its definition does not require $W^{ij}$ to be nowhere vanishing. 

There is a simple way to derive ${\bm W}^{i j}$ from $\bm G{}^{{i}{j}} $ in the case of  Minkowski superspace ${\mathbb M}^{3|8}$, for which the algebra of covariant derivatives  $D_A = (\pa_a, D^{i\bai}_\a)$ is obtained from \eqref{N=4MSS} by setting $X=0$. In ${\mathbb M}^{3|8}$, any constant real isotriplet is covariantly constant, so we may choose 
\bea
W_{\bai \baj} = \frac{\ri}{\sqrt{2} \k} (\s_1)_{\bai \baj} + \D W_{\bai\baj}~,
\label{W}
\eea
where $\k$ is a constant parameter, and  $\D W_{\bai \baj}$ is an infinitesimal right real linear superfield,
\bea
D_\a^{i(\bai}\D W^{\bak\bal)}=0~.
\eea
We now make use of the representation \eqref{W} to compute \eqref{G} in
${\mathbb M}^{3|8}$. Keeping only linear contributions in $\D W_{\bai \baj}$ leads to 
\bea
{\bm G}^{ij} = -\frac{\ri}{6\k}D^{ij\bai\baj}\D W_{\bai \baj}~.
\eea

Our model \eqref{super-ED} may be naturally generalised to the $\cN=4$ super Yang-Mills case, in a manifestly $\cN=4$ supersymmetric way, following the AdS construction of \cite{KTM14}. For simplicity, in the next section we will describe such a theory in terms of $\cN=2$ superfields on ${\mathbb M}^{3|4}_X$.

%%%%%%%%%%%%%%%%%%%%%%%%%%%%%%%%%%

\section{Topologically massive $\cN=4$ gauge theory in ${\mathbb M}^{3|4}_X$} 
\label{GaugeTheorySection}

In this section, we recast the actions for the left $\cN=4$ vector multiplet and hypermultiplet, which have been discussed earlier, in terms of $\cN=2$ superfields in ${\mathbb M}^{3|4}_X$. We will also describe super Yang-Mills couplings for these $\cN=4$ multiplets formulated in  ${\mathbb M}^{3|4}_X$. The actions for the right multiplets may be obtained from the left ones by applying the mirror map (\ref{MirrorMap}).

%%%%%%%%%%%%%%%%%%%%%%%%%%%%%%%%%%%%

\subsection{Gauge theory in ${\mathbb M}^{3|4}_X$}

So far we have worked with the covariant derivatives ${\cal D}_A = ({\cal D}_a , {\cal D}_\alpha,\bar{\cal D}^\alpha)$ for ${\mathbb M}^{3|4}_X$, which obey the graded commutations (\ref{N=2MSS}),  without giving any explicit realisation for these operators. Throughout this section, we make use of the following realisation  
\begin{equation}
    {\cal D}_\alpha = e^{-\hat{\cal V}}D_\alpha e^{\hat{\cal V}} = D_\alpha-2\ri X\bar\theta_\alpha \hat{\cal Z}\,,\qquad 
    \bar{\cal D}_\alpha = \bar D_\alpha\,,\qquad 
    \hat{\cal V} = -2\ri X \bar\theta^\alpha\theta_\alpha \hat{\cal Z}\,,
\label{D-relations}
\end{equation}
where $D_\alpha$ and $\bar D_\alpha$ are the spinor covariant derivatives for ${\mathbb M}^{3|4}$, the standard $\cN=2$ Minkowski superspace without central charge. The latter operators satisfy the algebra
\begin{equation}
    \{ D_\alpha , \bar D_\beta \} =-2\ri \partial_{\alpha\beta} = -2\ri {\cal D}_{\alpha\beta}\,,\qquad
    \{D_\alpha , D_\beta\} =0\,.
\end{equation}
The background superfield $\hat \cV$ in \eqref{D-relations} can be interpreted as the prepotential for a frozen $\sU(1)$ vector multiplet with a constant field strength (equal to $-2X\hat{\cal Z}$). Following the standard terminology of \cite{GGRS}, Eq.~\eqref{D-relations} defines the covariantly chiral representation of the covariant derivatives. 

Gauge theory in the $\cN=2$ superspace is described by gauge-covariant superspace derivatives $\nabla_A= (\nabla_a,\nabla_\alpha,\bar\nabla^\alpha)={\cal D}_A + \ri V_A$
obeying the following algebra
\begin{subequations}
\label{Dalgebra}
    \begin{align}
    \{ \nabla_\alpha,\bar\nabla_\beta \} =& -2\ri \nabla_{\alpha\beta} - 2\ri \varepsilon_{\alpha\beta}X\hat{\cal Z} + \ri \varepsilon_{\alpha\beta} G\,,\\
    [\nabla_\alpha, \nabla_a] =& -\frac12(\gamma_a)_\alpha{}^\beta\,  \nabla_{\beta}G \,,\qquad
    [\bar\nabla_\alpha, \nabla_{a}] = \frac12 (\gamma_a)_\alpha{}^\beta \,  \bar\nabla_{\beta}G \,,\\
    [\nabla_a,\nabla_b] =& -\frac\ri8 \varepsilon_{abc}(\gamma^c)^{\alpha\beta} [\nabla_\alpha,\bar\nabla_\beta]G\,,
    \label{algebra-F}
    \end{align}
\end{subequations}
with a covariantly linear superfield strengths $G$,
\begin{equation}
    \nabla^2 G= \bar\nabla^2 G =0\,.
    \label{linearity}
\end{equation}

The superfield strength may be expressed via a real prepotential $V=V^\dag$ transforming under gauge group by the rule
\begin{equation}
    \re^{V} \to \re^{\ri\lambda^\dag}\re^{V} \re^{-\ri\lambda}\,,
\end{equation}
where $\lambda$ is a chiral Lie-algebra-valued gauge parameter, $\bar D_\alpha \lambda = 0$. In this work, we will use the following asymmetric realization of the gauge-covariant superspace derivatives
\begin{equation}
    \nabla_\alpha = \re^{-V} D_\alpha \re^{V}\,,\qquad
    \bar\nabla_\alpha = \bar D_\alpha\,.
\label{GaugeCovDerivatives}
\end{equation}
Making use of the algebra (\ref{Dalgebra}), we derive the following expression for the superfield strength
\begin{equation}
    G = \frac \ri2 \bar D^\alpha (\re^{-V}D_\alpha \re^{V})\,.
\end{equation}
Note that the gauge prepotential $V$ is neutral with respect to the central charge, $\hat{\cal Z} V=0$.

The classical action of the topologically massive Yang-Mills theory in the $\cN=2$ superspace has the following form
\begin{equation}
    S^{\cN=2}_\mathrm{YM} = -\frac1{2g^2} \tr\int \mathrm{d}^{3|4}z
    \left[ G^2
    +\ri X\int_0^1 \mathrm{d}t\, \bar D^\alpha(\re^{-tV}D_\alpha \re^{tV})\re^{-tV}\partial_t \re^{tV}
    \right].
\label{N2SYM}
\end{equation}
Here $g$ is a coupling constant of mass dimension 1/2 whereas the last term represents the Chern-Simons action in the $\cN=2$ superspace found in Ref.~\cite{Ivanov:1991fn}. The  relative coefficient between the two terms in the action (\ref{N2SYM}) is not fixed by $\cN=2$ supersymmetry and can be arbitrary. We have fixed it to agree with the relevant  part of the $\cN=4$ super Yang-Mills Chern-Simons theory \eqref{MassiveSYM}.

%%%%%%%%%%%%%%%%%%%%%%%%%%%%%%%%%%

\subsection{$\cN=4$ super Yang-Mills Chern-Simons theory}

To describe the left $\cN=4$ super Yang-Mills (SYM) multiplet on ${\mathbb M}^{3|8}_X$, one follows the supergravity formalism of \cite{KTM14,KN14} and introduces gauge covariant derivatives
\bea 
{\mathfrak D}_A = \cD_A +\ri \,{\mathfrak V}_A ~,
\eea
where $\cD_A = (\cD_a, \cD^{i\bai}_\a)$ are the $\cN=4$ covariant derivatives for ${\mathbb M}^{3|8}_X$. The SYM connection ${\mathfrak V}_A $ is subject to constraints such that 
\bea
\{ {\frak D}_\a^{i\bai},{\mathfrak D}_\b^{j\baj }\}&=&
2\ri\,\ve^{ij}\ve^{\bai \baj }{\mathfrak D}_{\a\b}
+\,{2\ri}\ve_{\a\b}X (\ve^{\bai \baj } \bL^{ij} - \ve^{ij} \bR^{\bai\baj})
+2\ve_{\a\b} \ve^{i j } \mathfrak{W}^{\bai \baj}~.
\eea
Here the SYM field strength $\mathfrak{W}^{\bai \baj}$ is Hermitian, $(\mathfrak{W}^{\bai \baj})^\dagger = \mathfrak{W}_{\bai \baj}$, and obeys the Bianchi identity 
\bea
{\mathfrak D}^{i (\bar i}_\g \mathfrak{W}^{\baj \bar k)} =0~.
\eea

Upon reduction to the $\cN=2$ superspace ${\mathbb M}^{3|4}_X$,
$\mathfrak{W}^{\bai \baj}$ turns into three covariant superfields in the adjoint representation of the gauge group, which are: the $\cN=2$ SYM field strength 
$G$, a covariantly chiral scalar $\varPhi$ and its Hermitian conjugate $\bar \varPhi$,
\begin{equation}
    \nabla_\alpha\bar\varPhi = 0\,,\qquad
    \bar\nabla_\alpha\varPhi =0\,.
\end{equation}
Here $\nabla_A= (\nabla_a,\nabla_\alpha,\bar\nabla^\alpha)={\cal D}_A + \ri V_A$
are the $\cN=2$ gauge covariant derivatives. Introducing the Hermitian prepotential $V$ for the $\cN=2$ SYM multiplet and switching to the covariantly chiral representation, we obtain  
\begin{equation}
    \bar\varPhi = \re^{-V} \bar\varphi \re^{V}\,,\qquad
    \varPhi = \varphi\,,
\end{equation}
where $\varphi$ is a chiral scalar superfield, $\bar{\cal D}_\alpha \varphi =0$, of unit $ \hat{\cal Z}$ charge,
\begin{equation}
    \hat{\cal Z} \varphi = \varphi\,,\qquad
    \hat{\cal Z} \bar\varphi = - \bar\varphi\,.
\label{PhiCharge}
\end{equation}

The action for the left $\cN=4$ super Yang-Mills Chern-Simons theory is obtained by adding the $\bar\varPhi\varPhi$ term to the topologically massive $\cN=2$ SYM action (\ref{N2SYM}):
\begin{align}
    S^{\cN=4}_\mathrm{YMCS} =\frac1{g^2} \tr\int \mathrm{d}^{3|4}z
    \left[\bar\varPhi\varPhi -\frac12 G^2
    -\frac{\ri X}{2}\int_0^1 \mathrm{d}t\, \bar D^\alpha(\re^{-tV}D_\alpha \re^{tV})\re^{-tV}\partial_t \re^{tV}
    \right].
\label{MassiveSYM}
\end{align}
Being manifestly $\cN=2$ supersymmetric, this action respects also non-manifest $\cN=2$ supersymmetry transformations 
\begin{subequations}
\label{HiddenSUSY}
\begin{align}
    \Delta V &:=  \re^{-V}\delta \re^{V}
    = 2\ri(\bar\rho \varPhi - \rho \bar\varPhi)\,,\\
    \Delta \varPhi &:= \delta \varphi = \frac\ri{8X}\bar\nabla^2 (G {\cal D}^2\rho)\,, \label{DeltaPhi}\\
    \Delta\bar\varPhi &:= \re^{-V}\delta\bar\varphi \,\re^{V} = -\frac\ri{8X}\nabla^2(G\bar{\cal D}^2\bar\rho)\,,
    \label{DeltaBarPhi}
\end{align}
\end{subequations}
where we have denoted $\rho\equiv\rho_\mathrm{L}$. As is shown in section \ref{SecReduction}, the superfield parameter $\rho$ is 
 $x$-independent, covariantly chiral, and carries unit $\hat{\cal Z}$-charge,
\begin{equation}
    {\cal D}_a \rho =0\,,\qquad 
    \bar{\cal D}_\alpha \rho 
    = 0\,,\qquad
    \hat{\cal Z} \rho = \rho
  \,.
\label{UpsilonCharge}
\end{equation}
The invariance of the action (\ref{MassiveSYM}) under the transformations (\ref{HiddenSUSY}) may be verified using the properties (\ref{linearity}) and the following identities: 
\bea\bar {\cal D}^\alpha{\cal D}^2\rho = -4\ri X {\cal D}^\alpha\rho~, \qquad\bar{\cal D}^2{\cal D}^2 \rho = -16X^2 \rho~.
\eea
The above transformations represent a non-Abelian generalization of the variations (\ref{HiddenSusyAbelian}).

The action (\ref{MassiveSYM}) is invariant under the following gauge transformations:
\begin{equation}
    \varphi \to \re^{\ri\lambda}\varphi \,\re^{-\ri\lambda}\,,\quad
    \bar\varphi \to \re^{\ri\lambda^\dag} \bar\varphi\, \re^{-\ri\lambda^\dag} \,,\quad
    \re^{V} \to \re^{\ri\lambda^\dag} \re^{V} \re^{-\ri\lambda}\,.
\end{equation}
Here $\lambda$ is a chiral superfield taking its values in the Lie algebra of the gauge group, as discussed in the previous subsection. 

Note that the model (\ref{MassiveSYM}) arises in the flat-space limit from the corresponding manifestly $\cN=4$ supersymmetric model in AdS$_3$ \cite{KTM14}. The analogous model on a three-sphere was constructed in Refs.~\cite{HHL,SamsonovSorokin}. In the component field approach, $\cN=4$ super Yang-Mills Chern-Simons model was found in Ref.~\cite{LM}. In Appendix \ref{ComponentSYM} we present the component field form of the action (\ref{MassiveSYM}) together with the corresponding on-shell supersymmetry transformations.

In the Abelian case, the action (\ref{MassiveSYM}) simplifies to
\begin{equation}
    S_\mathrm{VM}^{\cN=4} 
    =\int \mathrm{d}^{3|4}z \left( \bar\varphi\varphi -\frac12 G^2 -\frac X2 VG\right)\,.
\label{Sabelian}
\end{equation}

It is well known that in three dimensions, no pure  supersymmetric Chern-Simons action exists for $\cN>3$, see e.g.\ the analysis in section 3 of \cite{KN14}. In the $\cN=4$ case, however, there exist uniquely defined supersymmetric Yang-Mills Chern-Simons actions for the following backgrounds: (i) the $(4,0)$ AdS superspace with $X\neq 0$ \cite{KTM14}; and (ii) the deformed Minkowski superspace.

%%%%%%%%%%%%%%%%%%%%%%%%%%%%%%%%%%%%%%%%%%%%%%

\subsection{Hypermultiplet in a gauge superfield background}

As is shown in Sec.~\ref{FShyper}, the Fayet-Sohnius hypermultiplet is described by a pair of chiral superfields $(Q_+,Q_-)$ with $\hat{\cal Z}$-charge $-\frac12$, 
\begin{equation}
    \bar{\cal D}_\alpha Q_\pm =0\,,\qquad
    \hat{\cal Z} Q_\pm =-\frac12 Q_\pm \,.
    \label{HyperCharges}
\end{equation}
For simplicity, in this section we will assume that these superfield transform in the fundamental and anti-fundamental representations of the gauge group,
\begin{equation}
    Q_+ \to \re^{\ri\lambda} Q_+\,,\qquad
    Q_- \to Q_-\re^{-\ri\lambda}\,.
\end{equation}
The generalization to any other representation is straightforward.

It is convenient to introduce covariantly chiral superfields,
\begin{equation}
    {\cal Q}_\pm = Q_\pm \,,\quad
    \bar{\cal Q}_+ = \bar Q_+ \re^{V}\,,\quad
    \bar{\cal Q}_- = \re^{-V}\bar Q_-\,,
    \label{CovQ}
\end{equation}
which are annihilated by the gauge-covariant superspace derivatives,
\begin{equation}
    \bar \nabla_\alpha {\cal Q}_\pm =0\,,\qquad
    \nabla_\alpha \bar{\cal Q}_\pm =0\,.
\end{equation}
In terms of these superfields, the hypermultiplet action reads
\begin{equation}
    S_\mathrm{hyper} = \int \mathrm{d}^{3|4}z \left[\bar{\cal Q}_+ {\cal Q}_+ + {\cal Q}_-\bar{\cal Q}_-\right] - \ri\int \mathrm{d}^{3|2}z\, {\cal Q}_- \varPhi {\cal Q}_+  
    +\ri\int \mathrm{d}^{3|2}\bar z\, \bar {\cal Q}_+ \bar\varPhi \bar {\cal Q}_-\,.
\label{ShyperQ}
\end{equation}
It is straightforward to check that this action is invariant under non-manifest $\cN=2$ supersymmetry transformations generated by the chiral superfield parameter $\rho$:
\begin{equation}
	\delta {\cal Q}_\pm = \pm \frac12 \bar\nabla^2(\bar\rho \bar {\cal Q}_\mp)\,,\qquad
	\delta\bar {\cal Q}_\pm = \pm \frac12\nabla^2 (\rho {\cal Q}_\mp)\,.
\label{deltaQnonabelian}
\end{equation}
Recall that the vector multiplet variations are given by Eqs.~(\ref{HiddenSUSY}). The variations (\ref{deltaQnonabelian}) represent a non-Abelian generalization of Eq.~(\ref{deltaQabeian}).

The chiral superfields $Q_\pm$ have the standard component field content: 
\begin{equation}
	Q_\pm| = f_\pm\,,\quad
	D^\alpha Q_\pm| = \chi_\pm^\alpha\,,\quad
	-\frac14 D^2 Q_\pm| = F_\pm\,.
\end{equation}
Focusing on the spinors $\chi_\pm^\alpha$, we find that the hypermultiplet action (\ref{ShyperQ}) contains the following terms among its components
\begin{equation}
    S_\mathrm{hyper} = -\frac\ri2 \int \mathrm{d}^3x\left[
	\chi_+^\alpha\left((\gamma^a)_{\alpha\beta}\nabla_a+\varepsilon_{\alpha\beta}X/2\right)\bar\chi_+^\beta
    + \chi_-^\alpha\left((\gamma^a)_{\alpha\beta}\nabla_a+\varepsilon_{\alpha\beta}X/2\right)\bar\chi_-^\beta
    \right] +\ldots
\label{SpinorComponents}
\end{equation}
Thus, both spinors $\chi_+^\alpha$ and $\chi_-^\alpha$ have the same sign in the mass terms, with mass $m=X/2$, but carry opposite charges with respect to the vector field, $\nabla_a\chi^\alpha_e = (\partial_a + \ri e A_a)\chi^\alpha_e$, $e=\pm$. As will be shown in the next section, this allows for the generation of the Chern-Simons term for the vector field in the effective action because the corresponding contributions from these spinors will not cancel out.

%%%%%%%%%%%%%%%%%%%%%%%%%%%%%%%%%%%%%%%%

\subsection{Topologically massive $\cN=4$ gauge theory from radiative corrections in the hypermultiplet model}

In this section, we consider the low-energy effective action of a hypermultiplet interacting with a background topologically massive $\cN=4$ vector multiplet. We will demonstrate that the action (\ref{Sabelian}) arises as a part of the hypermultiplet effective action. Here, for simplicity, we restrict ourselves to the Abelian case, although we expect that the same result holds for any non-Abelian gauge group.

In general, the effective action includes parity-odd and parity-even parts,
\begin{equation}
    \Gamma = \Gamma_\mathrm{odd}[V] + \Gamma_\mathrm{even}[V,\varphi]\,.
\end{equation}
The contribution to the parity-odd part can only have the form of the Chern-Simons term $\propto \int \mathrm{d}^{3|4}z \, VG$. Within the derivative expansion of the effective action, the parity-even part $\Gamma_\mathrm{even}$ may start from the Yang-Mills term for the $\cN=2$ vector multiplet, $\propto \int \mathrm{d}^{3|4}z\,G^2$, and its $\cN=4$ supersymmetric $\varphi$-completion, as well as should have higher-order derivative terms. Here we will be interested in only leading terms in the effective action and will ignore all terms with any derivatives of the superfield strength $G$ and the chiral superfield $\varphi$.

The formalism to perform perturbative calculations of low-energy effective actions in various three-dimensional gauge theories in $\cN=2$ superspace was developed in a series of papers \cite{BPS1,BPS2,BMS1,BMS2}. Here we will use this approach for calculations of the low-energy effective action of the hypermultiplet model (\ref{ShyperQ}) focusing on new features due to the presence of the non-trivial $\hat{\cal Z}$ central charge. For simplicity, we will consider the Abelian case where the hypermultiplet action (\ref{ShyperQ}) is part of the $\cN=4$ supersymmetric electrodynamics.

We start by considering the parity-odd part of the effective action represented by the Chern-Simons term. For deriving this term in the effective action it is sufficient to consider the gauge superfield background with vanishing chiral superfield, $\varphi=\bar\varphi=0$. On this background, the general variation of the effective action may be written as
\begin{equation}
    \delta \Gamma = \int \mathrm{d}^{3|4}z\,\delta V\langle J\rangle\,,
\end{equation}
where the effective current $\langle J\rangle $ reads
\begin{equation}
    \langle J\rangle = \langle \frac{\delta S_\mathrm{hyper}}{\delta V} \rangle = \langle {\cal Q}_+ \bar {\cal Q}_+\rangle -\langle  {\cal Q}_- \bar {\cal Q}_-\rangle = \sum_{e=\pm} e\langle {\cal Q}_e \bar {\cal Q}_e\rangle\,.
\label{Jeffective}
\end{equation}
The propagator for the chiral superfields may be formally written as
\begin{equation}
    \ri\langle {\cal Q}_e(z) \bar {\cal Q}_e(z') \rangle = \frac1{16}\frac 1{\hBox_e}\bar \nabla^2 \nabla'{}^2 \left[ \delta^{3|4}(z-z'){\cal I}(z,z')\right]\,,
\label{Qpropagator}
\end{equation}
where $\delta^{3|4}(z-z')$ is the full superspace delta function, ${\cal I}(z,z')$ is the parallel displacement operator,\footnote{By definition, this operator transforms  covariantly under the gauge group, reduces to the unit operator at coincident superspace points, ${\cal I}(z,z)= \bf 1$, and is annihilated by the operator $\zeta^A \nabla_A$, $\zeta^A \nabla_A {\cal I}(z,z')=0$, for a supersymmetric interval $\zeta^A$. More 
properties of the parallel displacement operator in three-dimensional gauge theories in the $\cN=2$ superspace are given in Ref.~\cite{BMS1}. See also Refs.~\cite{Kuzenko:2003eb,Kuzenko:2003qg,Kuzenko:2007cg} for applications of the parallel displacement operator to multiloop calculations of the effective action in four-dimensional supersymmetric gauge theories.} and $\hBox_e$ is the covariant box operator acting in the space of covariantly chiral superfields of charge $e$,
\begin{equation}
    \hBox_e{\cal Q}_e \equiv \frac1{16}\bar\nabla^2 \nabla^2{\cal Q}_e\,.
\end{equation}
Making use of the commutation relations (\ref{Dalgebra}), this operator may be cast in the form
\begin{equation}
	\hBox_e = \nabla^a \nabla_a - \frac{\ri e}2 W^\alpha \nabla_\alpha - \frac{\ri e}4  (\nabla^\alpha W_\alpha) - \left(-X\hat {\cal Z} + \frac{e}{2} G\right)^2\,,
	\label{CovBox}
\end{equation}
where $W_\alpha = \bar\nabla_\alpha G$.

The propagator (\ref{Qpropagator}) obeys equation $-\frac14\nabla^2 \ri \langle {\cal Q}_e(z) \bar {\cal Q}_e(z') \rangle = -\frac14 \nabla^2\left[ \delta^{3|4}(z-z') {\cal I}(z,z')\right]$, where the expression on the right-hand side plays the role of the delta-function in the space of covariantly antichiral superfields. The effective current (\ref{Jeffective}) is determined by this propagator in the limit of coincident superspace points.

We are interested in the leading terms in the effective current (\ref{Jeffective}) depending on the superfield $G$ without derivatives. For this purpose, it is sufficient to approximate the covariant box operator (\ref{CovBox}) by the following terms: $\hBox_e\approx \partial^a \partial_a - (-X\hat {\cal Z}+eG/2)^2$. In this approximation, all superspace derivatives in the propagator (\ref{Qpropagator}) should annihilate the Grassmann delta function due to the identity $\frac1{16}\bar{\cal D}^2 {\cal D}'{^2}\delta^4(\theta-\theta')|_{\theta=\theta'}=1$. Then, recalling that the $\hat{\cal Z}$-charge of the hypermultiplet is $-1/2$, see Eq.~(\ref{HyperCharges}), we can formally write $\hBox_e^{-1} \delta^3(x-x') = -\ri\int_0^\infty \mathrm{d}s\,\re^{\ri s[\partial^a \partial_a - \frac14(X + eG)^2]}\delta^3(x-x')$. This allows us to find the leading terms in the propagator at coincident superspace points,
\begin{equation}
    \langle {\cal Q}_e(z) \bar{\cal Q}_e(z) \rangle = \sqrt{-\ri} \int_0^\infty \frac{\mathrm{d}s}{(4\pi s)^{3/2}}\re^{-\ri\frac{s}{4}(X + e G)^2}
    =-\frac1{8\pi} |X + e G|\,,
\label{PropagatorResult}
\end{equation}
where $e=\pm$ is the charge of the chiral superfield with respect to the vector multiplet. Substituting Eq.~(\ref{PropagatorResult}) into (\ref{Jeffective}) we find the parity-odd part of the effective current:
\begin{equation}
    \langle J \rangle_\mathrm{odd} = -\frac1{8\pi}\left[(X + G) - (X - G)	\right] = -\frac{G}{4\pi}\,.
\end{equation}
As a result, we end up with the Chern-Simons term in the parity-odd part of the hypermultiplet effective action
\begin{equation}
    \Gamma_\mathrm{odd} = -\frac1{8\pi} \int \mathrm{d}^{3|4}z\,VG\,.
\label{GammaOdd}
\end{equation}

The action (\ref{GammaOdd}) has the unique $\cN=4$ completion with respect to the non-manifest $\cN=2$ supersymmetry transformations (\ref{HiddenSUSY}):
\begin{equation}
    \Gamma = \frac1{4\pi X}\int \mathrm{d}^{3|4}z \left(\bar\varphi\varphi -\frac12 G^2 - \frac X2 VG \right).
\end{equation}
For $X=g^2/(4\pi)$ this expression coincides with the classical action of topologically massive Abelian $\cN=4$ gauge theory (\ref{Sabelian}). The Maxwell term in this action $\int \mathrm{d}^{3|4}z\left(\bar\varphi\varphi -G^2/2\right)$ can be derived directly by expanding the one-loop effective action $\Gamma = \frac\ri2\sum_e \Tr \ln\hBox_e$ and keeping two-derivative terms in this expansion. This procedure, although technical, is straightforward, with relevant details described in Refs.~\cite{BPS1,BPS2,BMS1,BMS2}.

At first sight, it might seem that the presence of the Chern-Simons term (\ref{GammaOdd}) in the hypermultiplet effective action contradicts the results of the works \cite{Quantum3DHSS,BPS1}. Indeed, in these works, it was demonstrated that the Chern-Simons term cannot appear in the effective action of a massive hypermultiplet interacting with a background vector multiplet because the corresponding contributions from ${\cal Q}_+$ and ${\cal Q}_-$ cancel out. This is indeed the case when the central charge appears as a vev of the physical vector multiplet $(V,\varphi)$. In the present paper, we consider a different realization of the central charge in the $\cN=2$ superspace attributed to the superfield ${\cal V} = \ri\bar\theta^\alpha \theta_\alpha X$. In this realization, all physical spinors in the hypermultiplet have the same sign in the mass term, see Eq.~(\ref{SpinorComponents}). It is known (see, e.g., Refs.~\cite{Niemi:1983rq,Redlich1,Redlich2}) that a complex spinor of mass $m$ makes the following contribution to the Chern-Simons level (in the Abelian case)
\begin{equation}
	\frac1{8\pi}\frac{m}{|m|}\int \mathrm{d}^3x\,\varepsilon^{abc}A_a \partial_b A_c\,.
\end{equation}
Thus, the contributions from the spinors in Eq.~(\ref{SpinorComponents}) to the parity-odd part of the effective action add up in our case while these cancel out in the hypermultiplet model studied in Refs.~\cite{Quantum3DHSS,BPS1}.

We also stress that the Chern-Simons term in the effective action (\ref{GammaOdd}) is not renormalized beyond one loop.
\\

\noindent
{\bf Acknowledgements:}\\
GTM is grateful to Hongliang Jiang for discussions. The work of SMK and IBS is supported in part by the Australian Research Council, project No. DP230101629.
The work of ESNR is supported by the Brian Dunlop Physics Fellowship. 
The work of GTM was supported by the Australian Research Council (ARC) Future Fellowship FT180100353, ARC Discovery Project DP240101409, the Capacity Building Package at the University of Queensland and a faculty start-up funding of UQ's School of Mathematics and Physics.

%%%%%%%%%%%%%%%%%%%%%%%%%%%%%%%%%%%%%%%%%%%%%%%%%%

\appendix

\section{Notation and conventions}
\label{3Dconventions}

This appendix follows \cite{KLT-M11}.
Our conventions for spinors in three spacetime dimensions (3D) are compatible with the 4D two-component spinor formalism used in \cite{WB,Ideas}. As usual, 3D vector indices are labeled by values $m=0,\,1,\,2$.
Given the Pauli matrices $\vec{\s}=(\s_1,\s_2,\s_3)$, we choose the 3D gamma-matrices 
\begin{subequations}
\bea
&(\g_0 )_{\a  \b} = {\mathbbm 1}
~,~~~
(\g_1 )_{\a  \b} = \s_1
~,~~~
(\g_2 )_{\a  \b} = \s_3
 ~,\\
& (\g_m )^{\a  \b} = (\g_m)^{\b\a}
=\ve^{\a \g} \ve^{\b \d} (\g_m)_{\g \d} ~,
\eea
\end{subequations}
where the spinor indices are  raised and lowered using
the $\sSL(2,{\mathbb R})$ invariant tensors
\bea
\ve_{\a\b}=\left(\begin{array}{cc}0~&-1\\1~&0\end{array}\right)~,\qquad
\ve^{\a\b}=\left(\begin{array}{cc}0~&1\\-1~&0\end{array}\right)~,\qquad
\ve^{\a\g}\ve_{\g\b}=\d^\a_\b
\eea
as follows:
\bea
\psi^{\a}=\ve^{\a\b}\psi_\b~, \qquad \psi_{\a}=\ve_{\a\b}\psi^\b~.
\eea
By construction, the matrices $ (\g_m )_{\a  \b} $ and $ (\g_m )^{\a  \b} $ are {\it real} 
and symmetric.
For the matrices
\be
\g_m:=(\g_m)_\a{}^{\b}=\ve^{\b\g}(\g_m)_{\a\g}
\ee
the following relations hold
\bsubeq
\bea
&\{\g_m,\g_n\}=2\eta_{mn}{\mathbbm 1}~,
\\
&\g_m\g_n=\eta_{mn}{\mathbbm 1}+\ve_{mnp}\g^p~,
\eea
\esubeq
where the 3D Minkowski metric is $\eta_{mn}=\eta^{mn}={\rm diag}(-1,1,1)$, 
and the Levi-Civita tensor is normalized as $\ve_{012}=-\ve^{012}=-1$.
Some useful relations involving $\g$-matrices are 
\bsubeq
\bea
(\g^a)_{\a \b} (\g_a)_{\g \d} &=& 2\ve_{\a(\g}  \ve_{\d)\b} ~,
\\
\ve_{abc}(\g^b)_{\a\b}(\g^c)_{\g\d}&=&
\ve_{\g(\a}(\g_a)_{\b)\d}
+\ve_{\d(\a}(\g_a)_{\b)\g}
~,
\\
\tr[\g_a\g_b\g_{c}\g_d]&=&
2\eta_{ab}\eta_{cd}
-2\eta_{ac}\eta_{db}
+2\eta_{ad}\eta_{bc}
~.
\eea
\esubeq

Given a three-vector $V_m$, it can equivalently be realized as a symmetric spinor 
$V_{\a\b} =V_{\b \a}$.
The relationship between $V_m$ and $V_{\a \b}$ is as follows:
\bea
V_{\a\b}:=(\g^a)_{\a\b}V_a=V_{\b\a}~,\qquad
V_a=-\hf(\g_a)^{\a\b}V_{\a\b}~.
\label{vector-rule}
\eea
In three-dimensions an antisymmetric tensor $F_{ab}=-F_{ba}$ is Hodge-dual to a three-vector $F_a$, specifically
\bea
F_a=\hf\ve_{abc}F^{bc}~,\qquad
F_{ab}=-\ve_{abc}F^c~.
\label{hodge-1}
\eea
Then, the symmetric spinor $F_{\a\b} =F_{\b\a}$, which is associated with $F_a$, can equivalently be defined in terms of  $F_{ab}$: 
\bea
F_{\a\b}:=(\g^a)_{\a\b}F_a=\hf(\g^a)_{\a\b}\ve_{abc}F^{bc}
~.
\label{hodge-2}
\eea
These three algebraic objects, $F_a$, $F_{ab}$ and $F_{\a \b}$, are in one-to-one correspondence to each other, $F_a \leftrightarrow F_{ab} \leftrightarrow F_{\a\b}$. The corresponding inner products are related to each other as follows:
\bea
-F^aG_a=
\hf F^{ab}G_{ab}=\hf F^{\a\b}G_{\a\b}
~.
\eea

Let $\cM_{ab}=-\cM_{ba}$  be the $\sSL(2,{\mathbb R})$ generators. They act on a vector $V_a$ as 
\bea
\cM_{ab}V_c=2\eta_{c[a}V_{b]}~,
\eea
and on a spinor $\j_\a$ as  
\bea
\cM_{ab}\psi_\a
=\hf\ve_{abc}(\g^c)_\a{}^\b\psi_\b
~.
\eea
In accordance with
(\ref{vector-rule})--(\ref{hodge-2}), the Lorentz generators can also be realized as the vector $\cM_a$ or the symmetric spinor $\cM_{\a\b}$ such that 
\bea
\cM_{a}\psi_\a
=- \hf (\g_a)_\a{}^\b\psi_\b~, \qquad 
\cM_{\a\b}\psi_\g
=\ve_{\g(\a}\psi_{\b)}
~.
\eea
As is clear from the explicit form of the $\g$-matrices, we are using a Majorana representation in which all the $\g$-matrices are real, and any Majorana spinor $\j^\a$ is real,  
\bea
(\psi^{\a})^* = \psi^{\a} ~,\qquad (\psi_{\a})^*= \psi_{\a}  ~.
\eea

In this paper we often make use of the group isomorphisms  $\sSO(4) \cong  \big( {\sSU}(2)_{\rL}\times {\sSU}(2)_{\rR} \big)/{\mathbb Z}_2$ in order to convert each  $\sSO(4)$ vector index into a pair of $\sSU(2)$ ones. We first  introduce the following $\S$-matrices
\bea
(\S_I)_{i\bar{i}}=({\mathbbm 1},\ri\s_1,\ri\s_2,\ri\s_3)
~,~~~~~~
I={\bf 1},\cdots,{\bf 4}~,~~~
i=1,2~,~~
\bar{i}=\bar{1},\bar{2}
~.
\label{A.16}
\eea
The index $I$ is an $\sSO(4)$  vector one, while the indices $i$ and  $\bar{i}$  are, respectively, $\sSU(2)_{\rL}$ and $\sSU(2)_{\rR}$ spinor indices. Under complex conjugation the $\S$-matrices satisfy the reality property
\bea
\big((\S_I)_{i\bar{i}}\big)^*=(\S_I)^{i\bar{i}} =\ve^{ij} \ve^{\bar i \bar j}(\S_I)_{j\bar{j}}~.
\eea
Given $\sSU(2)_{\rL}$ and $\sSU(2)_{\rR}$ spinors $\psi_i$ and $\chi_{\bar{i}}$, respectively, we raise and lower their indices by using the antisymmetric tensors $\ve^{ij},\ve_{ij}$ and $\ve^{{\bar i}{\bar j}},\ve_{{\bar i}{\bar j}}$ ($\ve^{12}=\ve_{21}=\ve^{{\bar 1}{\bar 2}}=\ve_{{\bar 2}{\bar 1}}=1$) according to the rules:
\bea
\psi^{i}=\ve^{ij}\psi_j~,~~~
\psi_{i}=\ve_{ij}\psi^j
~,~~~~~~
\chi^{{\bar i}}=\ve^{\bai\baj}\chi_\baj~,~~~
\chi_{\bai}=\ve_{\bai\baj}\chi^\baj
~.
\eea
For practical calculations, it is useful to  introduce the $\t$-matrices
\bea
(\t_I)_{i\bar{i}}:=\frac{1}{\sqrt{2}}(\S_{I})_{i\bar{i}}
\eea
which have the following properties:
\bsubeq
\bea
(\t_{(I})_{i\bar{j}}(\t_{J)})^{j\bar{j}}=\hf\d_{IJ}\d_i^j
~&,&~~~
(\t_{(I})_{j\bar{i}}(\t_{J)})^{j\bar{j}}=\hf\d_{IJ}\d_{\bar{i}}^{\bar{j}}
~,
\\
(\t_I)_{i\bar{i}}(\t^I)_{j\bar{j}}=\ve_{ij}\ve_{\bar{i}\bar{j}}~&,&~~~
(\t_I)_{i\bar{i}}(\t_J)^{i\bar{i}}=\d_{IJ}
~.
\eea
\esubeq
It is the $\t$-matrices which are used in the paper to convert each SO(4) vector index  to a pair of isospinor ones,  $I\,\to\,i\bar{i}$. Associated with  an SO(4) vector $A_{I}$ is the second-rank isospinor $A_{i\bar{i}}$ defined by
\bea
A_{i\bar{i}}:=(\t_I)_{i\bar{i}}A^{I} \quad \longleftrightarrow \quad
A_{I}=(\t_I)^{i\bar{i}}A_{i\bar{i}}
~.
\eea
With the normalization chosen for the $\t$-matrices, it holds that 
\bea
\d_I^J\,\to\,\d_i^j\d_{\bar{i}}^{\bar{j}}
~,~~~~~~
A_{I}B^{I}= A_{i\bar{i}}B^{i\bar{i}}
~.
\eea

Given an antisymmetric second-rank $\sSO(4)$ tensor, $A_{IJ}=-A_{JI}$, its counterpart with isospinor indices, $A_{i\bai j\baj}=-A_{j\baj i\bai}=A_{IJ}(\t^I)_{i\bai}(\t^J)_{j\baj}$ can be decomposed as
\bea
A_{i\bai j\baj}=\ve_{ij}A_{\bai\baj}+\ve_{\bai\baj}A_{ij}~ \longrightarrow ~
A^{i\bai j\baj}=-\ve^{ij}A^{\bai\baj}-\ve^{\bai\baj}A^{ij}~,\qquad
A_{ij}=A_{ji}~,~~A_{\bai\baj}=A_{\baj\bai}
~.~~~
\eea
Here the two independent symmetric isospinors  $A_{ij}$ and $A_{\bai\baj}$ represent the self-dual and anti-self-dual parts of the antisymmetric tensor $A_{IJ}$. Given another antisymmetric second-rank $\sSO(4)$ tensor, 
$B_{IJ}=-B_{JI}$,
and the corresponding isospinor counterparts 
$B_{ij} $ and $B_{\bai\baj}$, one can check that 
\bea
\hf A^{IJ}B_{IJ}
&=& A^{ij}B_{ij}+A^{\bai\baj}B_{\bai\baj}
~.
\eea
Finally, consider the completely antisymmetric fourth-rank tensor $\ve_{IJKL}$, which is normalized by 
$\ve_{{\bf 1}{\bf 2}{\bf 3}{\bf 4}}=1$.
Its isospinor counterpart is
\bea
\ve_{i\bai j\baj k\bak l\bal}&:=&
\ve_{IJKL}(\t^I)_{i\bai}(\t^J)_{j\baj}(\t^K)_{k\bak}(\t^L)_{l\bal} 
=\Big(\ve_{ij}\ve_{kl}\ve_{\bai\bal}\ve_{\baj\bak}
-\ve_{il}\ve_{jk}\ve_{\bai\baj}\ve_{\bak\bal}\Big)
~.
\eea

\section{(Hyper) K\"ahler cones} \label{Cone}

This appendix is taken verbatim from \cite{BKTM}.

Consider a K\"ahler manifold $(\cM, g_{\m\n}, J^\m{}_\n )$, where $\m,\n=1,\dots, 2n$,  
and introduce local complex coordinates
$\f^\ra$ and their conjugates $\bar \f^{\bar \ra}$, in which the complex structure 
$J^\m{}_\n$ is diagonal.
It is called a K\"ahler cone \cite{GibbonsR} if it possesses
a homothetic conformal Killing vector $\c$
 \bea 
\c = \c^\ra \frac{\pa}{\pa \f^\ra} + {\bar \c}^{\bar \ra}  \frac{\pa}{\pa {\bar \f}^{\bar \ra}}
\equiv \c^\m \frac{\pa}{\pa \vf^\m} 
\eea
which is the gradient of a function. These conditions mean that 
\bea
\tsD_\n \c^\m = \d_\n{}^\m \quad \Longleftrightarrow \quad 
\tsD_\rb \c^\ra= \d_\rb{}^\ra~, \qquad 
\tsD_{\bar \rb} \c^\ra = \pa_{\bar \rb} \c^\ra = 0~.
\label{hcKv}
\eea
In particular,  $\c $ is holomorphic. The properties of $\c$ include the following:
\bea
\c_\ra := {g}_{\ra \bar \rb} \,{\bar \c}^{\bar \rb} = \pa_\ra { K}\quad \Longrightarrow \quad 
\c^\ra K_\ra = K~, 
\label{hcKv-pot}
\eea
where 
\bea
K:={ g}_{\ra \bar \rb} \, \c^\ra {\bar \c}^{\bar \rb}
\label{hcKv-pot2}
\eea
can be used as a  K\"ahler potential, $ {g}_{\ra \bar \rb} = \pa_\ra \pa_{\bar \rb} K$.
Associated with $\c$ is the U(1) Killing vector field 
\bea 
V^\m = -\hf J^\m{}_\n \c^\n~, \qquad \nabla_\m V_\n + \nabla_\n V_\m =0~.
\eea

Local complex coordinates for $\cM$ can always be
chosen such that 
 \bea
\c = \f^\ra \frac{\pa}{\pa \f^\ra} + {\bar \f}^{\bar \ra}  \frac{\pa}{\pa {\bar \f}^{\bar \ra}}~, 
\eea
and then the second relation in \eqref{hcKv-pot} turns into the homogeneity condition 
\bea
\f^\ra K_\ra (\f, \bar \f)= K(\f, \bar \f)~,
\eea
compare with Eq. \eqref{Kkahler}.

A hyperk\"ahler manifold $(\cM, g_{\m\n}, (J_A)^\m{}_\n )$, where $\m,\n=1,\dots, 4n$
and $A=1,2,3$, 
is called a hyperk\"ahler cone \cite{deWRV}
if it is a K\"ahler cone with respect to each complex structure. 
Using $J_A$ and $\chi$, we may construct three $\rm SU(2)$ Killing vectors
\begin{align}
V_A^\mu := -\frac{1}{2} (J_A)^\mu{}_\nu \chi^\nu~.
\end{align}
These vectors commute with $\c$  and obey an $\rm SU(2)$ algebra amongst themselves,
\begin{align}
[V_A, \chi] = 0~, \qquad [V_A, V_B] = \epsilon_{ABC} V_C~.
\end{align}
They generate a transitive  action of SO(3)  on the two-sphere of complex structures, 
\begin{align}
\cL_{V_A} J_B = \epsilon_{ABC} J_C~.
\end{align}
More information about the hyperk\"ahler cones can be found in \cite{deWRV}.

%%%%%%%%%%%%%%%%%%%%%%%%%%%%%%%%%%%%%%%%%%%%%%%%%%%

\section{Left $\cN=4$ super Yang-Mills Chern-Simons action in component fields}
\label{ComponentSYM}

In this appendix we present the left $\cN=4$ super Yang-Mills Chern-Simons action \eqref{MassiveSYM} in component fields.

The $\cN=2$ vector multiplet $V$ contains a real scalar field $\sigma$, a vector $A_m$, an auxiliary field $D$, and spinors $\lambda_\alpha$, $\bar\lambda_\alpha=(\lambda_\alpha)^\dag$. To derive these components, it is customary to impose the Wess-Zumino gauge,
\begin{equation}
	V| =0 \,,\quad{\cal D}_\alpha V| = \bar{\cal D}_\alpha V| =0\,,\quad
	{\cal D}^2 V| = \bar{\cal D}^2 V|=0\,,
\end{equation}
where $|$ denotes the projection $\theta_\alpha = \bar\theta_\alpha =0$. The physical fields are defined by the following relations:
\begin{align}
	\frac12[D_\alpha,\bar D_\beta]V| &= (\gamma^m)_{\alpha\beta} A_m + \ri\varepsilon_{\alpha\beta} \sigma\,,\\
	\frac14 D^2 \bar D_\alpha V| &= \ri\bar\lambda_\alpha\,,\qquad \frac14 \bar D^2 D_\alpha V| = -\ri\lambda_\alpha\,, 
	\\
	-\frac18 \{ D^2, \bar D^2 \} V| &= D\,.
\end{align}

The chiral superfield $\varphi$ has the standard component field content: a complex scalar $f$, a spinor $\psi_\alpha$ (and its conjugate $\bar\psi_\alpha=(\psi_\alpha)^\dag$), and a complex auxiliary field $F$:
\begin{equation}
	\varphi| = \frac{\ri}{\sqrt2}f \,,\quad
	 D_\alpha \varphi| = \frac{\ri}{\sqrt2}\psi_\alpha\,,\quad
	-\frac14 D^2 \varphi| = \frac{\ri}{\sqrt2}F\,.
\end{equation}
The complex scalar field from this multiplet may be combined with the real scalar $\sigma$ from the vector multiplet into a $\sSO(3)\simeq \sSU(2)_\mathrm{L}$ triplet,
\begin{equation}
	\phi_\mathrm{I} =(\phi_1,\phi_2,\sigma) = \left( \frac{f+\bar f}{\sqrt2} , \ri\frac{f-\bar f}{\sqrt2},\sigma \right)
    ~,~~~~~~\mathrm{I}=1,2,3
    ~.
\end{equation}
The spinor components from these multiplets may also be re-arranged into a $\sSU(2)_\mathrm{L}$ doublet field $\psi_{i\alpha}$,
\begin{equation}
	\bar\psi^{1\alpha} = \bar\lambda^\alpha\,,\quad
	\psi_{1\alpha} = \lambda_\alpha\,,\quad
	\psi_{2\alpha} = \frac1{\sqrt2}\bar\psi_\alpha\,,\quad
	\bar\psi^{2\alpha} = \frac1{\sqrt2} \psi^\alpha\,.
\end{equation}
As a result, after elimination of the auxiliary fields, we find the following component field content of the action (\ref{MassiveSYM}):
\begin{subequations}
	\label{Scomponent}
	\begin{align}
		S^{\cN=4}_\mathrm{YM} &= \frac1{g^2} \tr\int \mathrm{d}^3x({\cal L}_\mathrm{vector} + {\cal L}_\mathrm{scalar} + {\cal L}_\mathrm{spinor})\,,\\
		{\cal L}_\mathrm{vector} &= -\frac14 F^{mn}F_{mn} + \frac{X}{2}\varepsilon^{mnp}(A_m \partial_n A_p + \frac23A_m A_n A_p)\,,\\
		{\cal L}_\mathrm{scalar} &=\frac12\phi^\mathrm{I} (\nabla^m \nabla_m - X^2) \phi_\mathrm{I} - \frac \ri2X\varepsilon^\mathrm{IJK}\phi_\mathrm{I}[\phi_\mathrm{J},\phi_\mathrm{K}] + \frac14[\phi^\mathrm{I},\phi^\mathrm{J}][\phi_\mathrm{I},\phi_\mathrm{J}]\,,\\
		{\cal L}_\mathrm{spinor} &=\ri\bar\psi^{i\alpha} ((\gamma^m)_\alpha{}^\beta\nabla_m + X\delta_\alpha^\beta )\psi_{i\beta} + \ri(\sigma^\mathrm{I})_i{}^j \bar\psi^{i\alpha}[\phi_\mathrm{I},\psi_{j\alpha}]\,.
	\end{align}
\end{subequations}
Here $F_{mn} = \partial_m A_n - \partial_n A_m + \ri [A_m,A_n]\propto
\varepsilon_{mnp}(\gamma^p)^{\alpha\beta} [\nabla_\alpha,\bar\nabla_\beta]G|_{\theta=0}$ is the Yang-Mills field strength, and $(\sigma^\mathrm{I})_i{}^j$ are $\sSO(3)\sim \sSU(2)_\mathrm{L}$ $\sigma$-matrices that satisfy the following equations
\begin{equation}
(\sigma^\mathrm{I})_j{}^i(\sigma^\mathrm{J})_i{}^k = \ri\varepsilon^\mathrm{IJK}(\sigma_\mathrm{K})_j{}^k + \delta^\mathrm{IJ}\delta^k_i\,,\qquad
	(\sigma^\mathrm{I})_{ij}(\sigma_\mathrm{I})^{kl} = -(\delta_i^k\delta_j^l+\delta_i^l\delta_j^k)\,.
\end{equation}

It is straightforward to verify that the action (\ref{Scomponent}) is invariant under the following $\cN=4$ supersymmetry variations of the fields:
\begin{subequations}
	\begin{align}
		\delta A_m &= \ri(\gamma_m)_\alpha{}^\beta (\bar\epsilon^{i\alpha}\psi_{i\beta} + \epsilon_{i\beta}\bar\psi^{i\alpha})\,,\\
		\delta\phi^\mathrm{I} &= (\sigma^\mathrm{I})^i{}_j(\bar\epsilon^{j\alpha}\psi_{i\alpha}+\epsilon_{i\alpha}\bar\psi^{j\alpha})\,,\\
		\delta\bar\psi^{i\alpha} &= -\frac 12 \varepsilon^{mnp}(\gamma_p)^\alpha{}_\beta \bar\epsilon^{i\beta}F_{mn} - \ri(\gamma^m)^\alpha{}_\beta (\sigma^\mathrm{I})^i{}_j \nabla_m \phi_\mathrm{I} \bar\epsilon^{j\beta} \nonumber\\& -\ri X \phi^\mathrm{I} (\sigma_\mathrm{I})^i{}_j \bar\epsilon^{j\alpha} + \frac12 \varepsilon^\mathrm{IJK} (\sigma_\mathrm{K})^i{}_j[\phi_\mathrm{I},\phi_\mathrm{J}]\bar\epsilon^{j\alpha}\,,\\
		\delta\psi_{i\alpha}&=  \frac 12 \varepsilon^{mnp}(\gamma_p)_\alpha{}^\beta \epsilon_{i\beta}F_{mn} - \ri(\gamma^m)_\alpha{}^\beta (\sigma^\mathrm{I})_i{}^j \nabla_m \phi^\mathrm{I} \epsilon_{j\beta} \nonumber\\& + \ri X\phi^\mathrm{I} (\sigma_\mathrm{I})_i{}^j \epsilon_{j\alpha} - \frac12 \varepsilon^\mathrm{IJK}(\sigma_\mathrm{K})_i{}^j [\phi_\mathrm{I},\phi_\mathrm{J}]\epsilon_{j\alpha}\,.
	\end{align}
\end{subequations}
These transformations close on shell as the auxiliary fields have been eliminated.

The action (\ref{Scomponent}) may be obtained as the flat-space limit of the corresponding model on the three-sphere \cite{HHL,SamsonovSorokin}  upon transitioning to Lorentz signature.

%%%%%%%%%%%%%%%%%%%%%%%%%%%%%%%%%%%%%%%%%%%%%%%%%%%

\begin{footnotesize}

\end{footnotesize}

\end{document}